\documentclass[
 reprint,
 amsmath,amssymb,
 aps,
]{revtex4-2}

\usepackage{graphicx}
\usepackage{bm}
\usepackage{physics}
\usepackage{xcolor}
\usepackage{ulem}
\usepackage{blkarray,bigstrut}
\usepackage{amsmath}
\usepackage{wrapfig}
\usepackage{hyperref}

\begin{document}

\title{Improved Heralded Single-Photon Source with a Photon-Number-Resolving Superconducting Nanowire Detector}

\author{Samantha I. Davis$^{1,2}$}
\author{Andrew Mueller$^{2,3}$}
\author{Raju Valivarthi$^{1,2}$}
\author{Nikolai Lauk$^{1,2}$}
\author{Lautaro~Narvaez$^{1,2}$}
\author{Boris Korzh$^4$}
\author{Andrew D. Beyer$^4$}
\author{Olmo Cerri$^1$} 
\author{Marco Colangelo$^5$} 
\author{Karl~K.~Berggren$^5$}
\author{Matthew D. Shaw$^4$}
\author{Si Xie$^{1,2,6}$}
\author{Neil Sinclair$^{1,2,7}$}
\author{Maria Spiropulu$^{1,2,*}$}

\affiliation{$^1$Division of Physics, Mathematics and Astronomy, California Institute of Technology, 1200 E California Blvd., Pasadena, CA 91125, USA}
\affiliation{$^2$Alliance for Quantum Technologies (AQT), California Institute of Technology, 1200 E California Blvd., Pasadena, CA 91125, USA}
\affiliation{$^3$Division of Engineering and Applied Science, California Institute of Technology, 1200 E California Blvd., Pasadena, CA 91125, USA}
\affiliation{$^4$Jet Propulsion Laboratory, California Institute of Technology, 4800 Oak Grove Dr., Pasadena, CA 91109, USA}
\affiliation{$^5$Department of Electrical Engineering and Computer Science, Massachusetts Institute of Technology, 50 Vassar St., Cambridge, MA 02139, USA}
\affiliation{$^6$Fermi National Accelerator Laboratory, P.O. Box 500, Batavia, IL 60510, USA}
\affiliation{$^7$John A. Paulson School of Engineering and Applied Sciences, Harvard University, 29 Oxford St., Cambridge, MA 02138, USA}

\affiliation{$^*$Corresponding author: smaria@caltech.edu}

%\date{\today}

\begin{abstract}
Deterministic generation of single photons is essential for many quantum information technologies. 
A bulk optical nonlinearity emitting a photon pair, where the measurement of one of the photons heralds the presence of the other, is commonly used with the caveat that the single-photon emission rate is constrained due a tradeoff between multi-photon events and pair emission rate.  
Using an efficient and low noise photon-number-resolving superconducting nanowire detector we herald, in real time, a single photon at telecommunication wavelength.  
We perform a second-order photon correlation $g^2(0)$ measurement of the signal mode conditioned on the measured photon number of the idler mode for various pump powers and demonstrate an improvement of a heralded single photon source. We develop an analytical model using a phase space formalism that encompasses all multi-photon effects and relevant imperfections, such as loss and multiple Schmidt modes. 
We perform a maximum likelihood fit to test the agreement of the model to the data and extract the best-fit mean photon number $\mu$ of the pair source for each pump power.  
A maximum reduction of $0.118\pm0.012$ in the photon $g^2(0)$ correlation function at $\mu=0.327\pm0.007$ is obtained, indicating a strong suppression of multi-photon emissions.  
For a fixed $g^2(0)=7\times 10^{-3}$, we increase the single pair generation probability by 25\%. 
Our experiment, built using fiber-coupled and off-the-shelf components, delineates a path to engineering ideal sources of single photons.
\end{abstract}

\maketitle

\begin{figure*}[ht!]
    \centering
    \includegraphics[width = \textwidth]{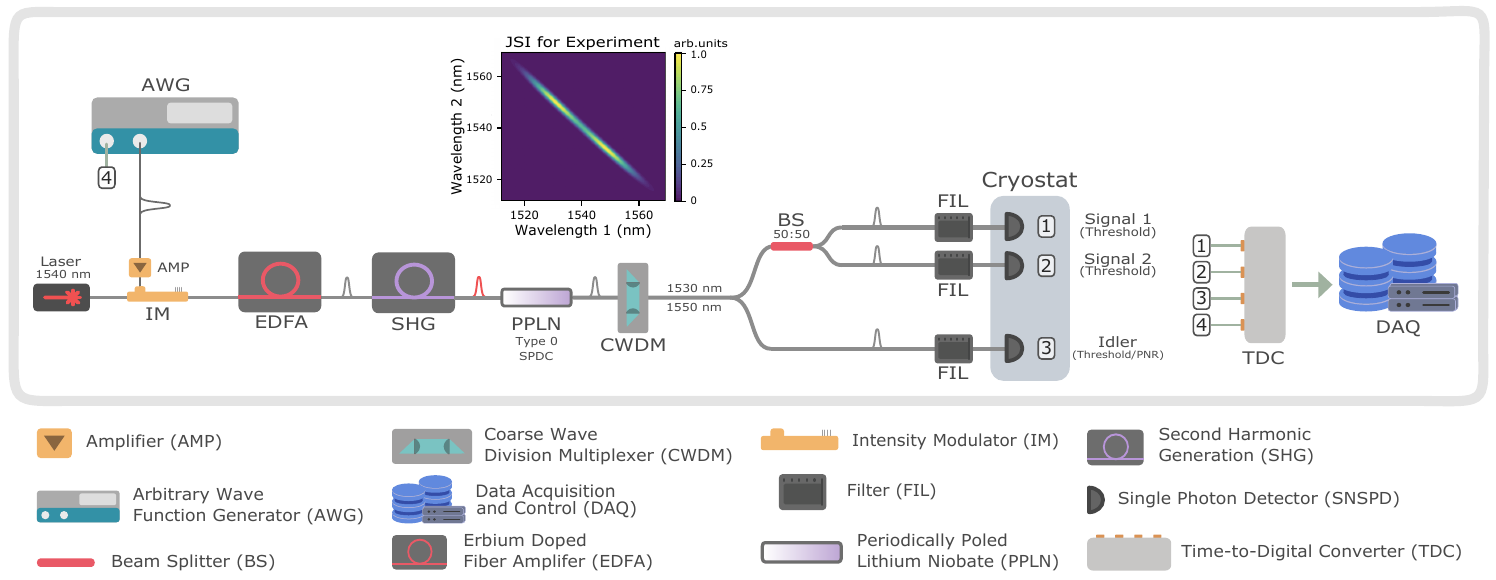}
    \caption{Experimental setup. AWG - Tektronix AWG7002A, BS - Thorlabs 1550nm fiber optic 50:50 beamsplitter, CWDM - FS one-channel coarse wave division multiplexing/optical add/drop multiplexer, EDFA - Pritel erbium-doped fiber amplifer,  Laser -  General Photonics TLS-101, PPLN - Covesion ruggedized waveguide, SHG - Pritel optical fiber amplifier/second harmonic generator. The inset shows the estimated joint spectral intensity (JSI) for the experiment including the detector and CWDM response.
    }
    \label{fig:setup}
\end{figure*}

\section{Introduction}
\label{sec:introduction}
A challenge in quantum optical science and technology is the realization of an ideal, i.e. deterministic, high-fidelity, tunable, and high-rate, source of indistinguishable single-photons \cite{eisaman2011invited,o2009photonic}.
One intuitive approach to develop a single-photon source requires coupling an individual quantum emitter to light using a cavity. Significant progress in this regard~\cite{aharonovich2016solid} has been achieved using e.g.  quantum dots ~\cite{michler2000quantum,shields2007semiconductor,senellart2017high}, crystal defects~\cite{babinec2010diamond}, or trapped ions~\cite{barros2009deterministic} and atoms ~\cite{mucke2013generation}, albeit mired with challenges, including fabrication complexity~\cite{bradac2019quantum,bogdanovic2017robust} or differing emitter spectra~\cite{huber2017highly,sipahigil2014indistinguishable,bernien2012two}. 
Instead, the strong light-matter coupling offered by solid-state bulk nonlinearities can  be used to probabilistically emit photon pairs via laser-driven $\chi^{(2)}$ and $\chi^{(3)}$ processes \cite{boyd2020nonlinear}, i.e. spontaneous parametric down-conversion (SPDC) and four-wave mixing (SFWM), respectively. Thermal statistics of the emission restrict a single photon pair to be emitted with low probability (e.g. $10^{-3}$ in practice \cite{valivarthi2020teleportation}).
An individual photon (in a signal mode) can be heralded by the detection of the other photon (in an idler mode) \cite{eisaman2011invited}. 
Typically this is performed using a threshold detector that discriminates zero from one or more photons. 
Heralding of photons from optical nonlinearities is scalable, and has enabled tunable and indistinguishable photons with high fidelities and bandwidths \cite{wang2020integrated,eisaman2011invited,spring2017chip}. However, there is a non-zero probability to produce multiple pairs. 
To overcome this obstacle, a photon number resolving (PNR) detector at the idler mode can be used to exclude multi-photon events. 
Notable demonstrations of PNR detection have used, e.g., transition edge sensors and pseudo-PNR detectors constructed from time-multiplexed or arrays of threshold detectors~\cite{lita2008counting,jiang2007photon,fitch2003photon}. 
Optimized heralded single photon sources require scalable, efficient, and low-noise PNR detectors with high timing resolution, that is, low jitter. 
Here we detect the idler mode from an SPDC process \textit{in real time} using a PNR niobium nitride (NbN) superconducting nanowire single-photon detector (SNSPD) \cite{colangelo2021}. 
The detector is optimized across several performance metrics \cite{natarajan2012superconducting}. 
Specifically, the detection efficiency, which includes coupling loss in the cryostat, is $>0.7$, the dark count rate is $10$ Hz, and the jitter is $<14$ ps. 

To quantify the improvement of our heralded single-photon source, we perform a second-order correlation function $g^2(0)$ measurement \cite{glauber1963quantum} of the signal mode conditioned on the measured photon number of the idler mode using the number-resolving detector. 
This measurement is performed as a function of mean photon-pair number $\mu$ of the source.

We operate the detector in two configurations: (i) as a PNR SNSPD, discriminating zero-, one- and multi-photon events, and (ii) as a threshold SNSPD, discriminating zero-photon events from all other events. 
A $g^2(0)$ of zero is expected when a single photon pair is detected. Accounting for loss and multi-photon events, a reduction in $g^2(0)$ is expected when the detector is operated in configuration (i) versus (ii) for a fixed $\mu$.
 
Since the measurements extend to large $\mu$, we develop an analytical model for the detection rates, coincidence rates, and $g^2(0)$ using a phase-space formalism that encompasses full multi-photon contributions and all relevant imperfections, such as loss and multiple Schmidt modes~\cite{takeoka2015full,feito2009measuring,achilles2004photon}. We model the PNR detector in phase space as a $2N$-port beamsplitter followed by threshold detection at each output, which allows us to employ Gaussian characteristic function techniques.  
To evaluate the single photon discrimination capability of the detector, we define the single photon discrimination efficiency $\eta_{PNR}^1$ metric, ranging from zero, for a threshold detector, to one for an ideal PNR detector.  
We obtain $\eta_{PNR}^1=0.46$ corresponding to a pseudo-PNR detector comprised of no more than 18 threshold detectors, each with efficiency $\eta_d=0.71$.
We perform a simultaneous maximum likelihood fit of the model to the measured values of $g^2(0)$ and extract $\mu$ for each pump power. 
We measure a maximum reduction of $g^2(0)$ from $0.430\pm0.009$ to $0.312\pm0.008$ when using configuration (ii) versus (i) at $\mu=0.327\pm0.007$, thereby improving the fidelity of the single photon source. 
For a fixed $g^2(0)=7\times 10^{-3}$~\cite{kaneda2019high}, we increase the probability to generate a single pair by 25\%, from $4\times 10^{-3}$ to $5\times 10^{-3}$.

\section{Experimental Methods}
\label{sec:experimentalmethods}
The experimental setup is shown in Fig. \ref{fig:setup}. Light pulses of $\sim600$~ps duration are created by injecting 1540~nm wavelength light from a continuous-wave laser into an intensity modulator (IM). 
The modulator is driven by an arbitrary waveform generator (AWG) at a rate $R=1$ MHz, which is the clock rate of the experiment. 
The pulses are amplified by an erbium doped fiber amplifier and then directed to a second harmonic generation module with a gain-adjusted amplification stage (SHG), which amplifies the pulses then up-converts them to 770 nm wavelength.
The pulses are then directed to a fiber-coupled type-0 periodically poled lithium niobate (PPLN) waveguide, which produces photon pairs centered at 1540 nm wavelength via SPDC. 
A coarse wavelength division multiplexer (CWDM) splits the photon pairs into the signal and idler modes, centered at 1530 nm and 1550 nm, respectively, each with a 13 nm bandwidth. 
Light in the signal path is split by a 50:50 beamsplitter (BS) into two paths, labelled as signal 1 and 2.  
Filters with a total of 60 dB extinction on the idler path and 120 dB extinction on the signal path are used to suppress the unconverted 770 nm pump light. 
The photons from the signal and idler paths are detected using conventional and PNR SNSPDs, respectively.

\subsection{Detectors}
\label{sec:detectors}
\begin{figure}[!h]
    \centering
    \includegraphics[width=\columnwidth]{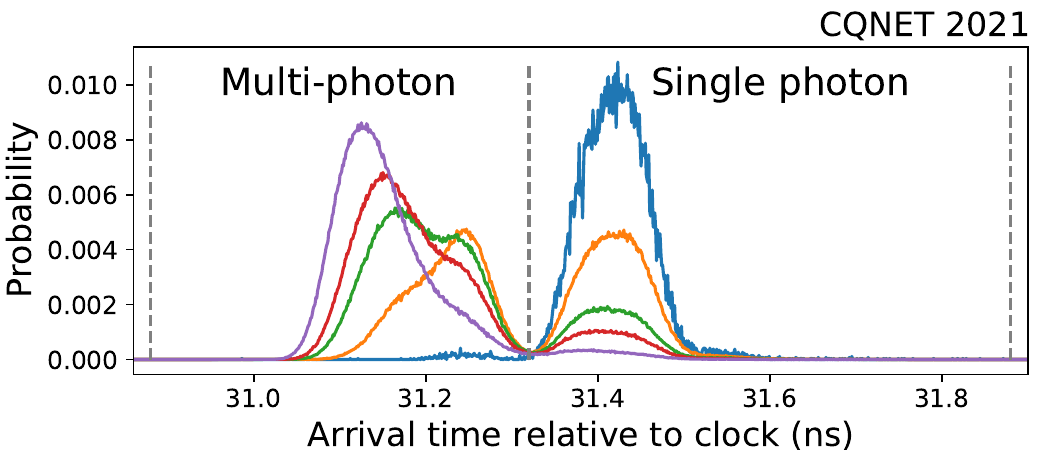}
    \caption{Probability distribution of the arrival times of detection events by the PNR SNSPD for $\mu \approx 8\times10^{-3}$ (blue), $\mu \approx 3$ (orange), $\mu \approx 9$ (green), $\mu \approx 11$ (red), and $\mu \approx 16$ (purple). The dashed lines define the time bins corresponding to single- (right) and multi-photon (left) events. The total number of events in the single- and multi-photon bins are used when operating the SNSPD as a threshold detector, while the number of events in the single-photon bin are used when operating the SNSPD as a PNR detector.}
    \label{fig:idler_pnr}
\end{figure}

The detectors are held at 0.8 K in a Gifford-McMahon cryostat with a $^4$He sorption stage.
To measure the signal modes, we use two single-pixel tungsten silicide (WSi) SNSPDs, which have timing jitters of $\sim 50$~ps, detection efficiencies of $\sim 0.8$, and dark count rates below 5~Hz~\cite{valivarthi2020teleportation}.
To measure the idler mode, we use a PNR SNSPD with a timing jitter of $< 14$ ps, detection efficiency of $\eta_d=0.71$ and dark counts $<10$ Hz. 
The detector efficiency was determined in an independent measurement similar to that performed in Ref.~\cite{marsili2013detecting}.
The detector has an active area of 22$\times$15 $\mu$m$^2$, formed by a meander of 100$~$nm-wide and 5$~$nm-thick NbN nanowires with a 500~nm pitch. 
The detector employs a differential architecture to cancel the contribution of the signal propagation delays to the timing jitter \cite{colangelo2021}.
An impedance-matching taper enables photon-number resolution, increases the signal-to-noise ratio, and minimizes reflections as well as distortion \cite{colangelo2021,zhu2019superconducting}.
The number of incident photons is encoded into the amplitude of the output pulse \cite{colangelo2021, zhu2020}.
A single incident photon that is absorbed by the nanowire induces a single time-dependent resistive hotspot, which results in a radio-frequency pulse \cite{natarajan2012superconducting}.
Multiple incident photons absorbed by the nanowire at the same time induce multiple time-dependent resistive hotspots. 
This increases the total resistance of the nanowire, producing a radio-frequency pulse with an amplitude and slew rate that depends on the number of hotspots. 
In our experiments, rather than measuring the pulse amplitude variation \cite{zhu2020,colangelo2021}, we measure its slew rate variation \cite{Cahall17}. 
This only requires a constant-threshold time tagger, i.e. time-to-digital converter, and enables real-time readout. 
With a fixed voltage threshold, the variation in slew rate results in a variation of the time of the detection event, i.e. time tag.
Earlier (later) time-tags, plotted in a histogram in the left (right) bin of Fig. \ref{fig:idler_pnr}, correspond to multi-photon (single-photon) pulses with higher (lower) slew rate.

\subsection{Data acquisition and analysis}
\label{sec:dataacquisitionandanalysis}
\begin{figure}[h!]
    \centering
    \includegraphics[width=\columnwidth]{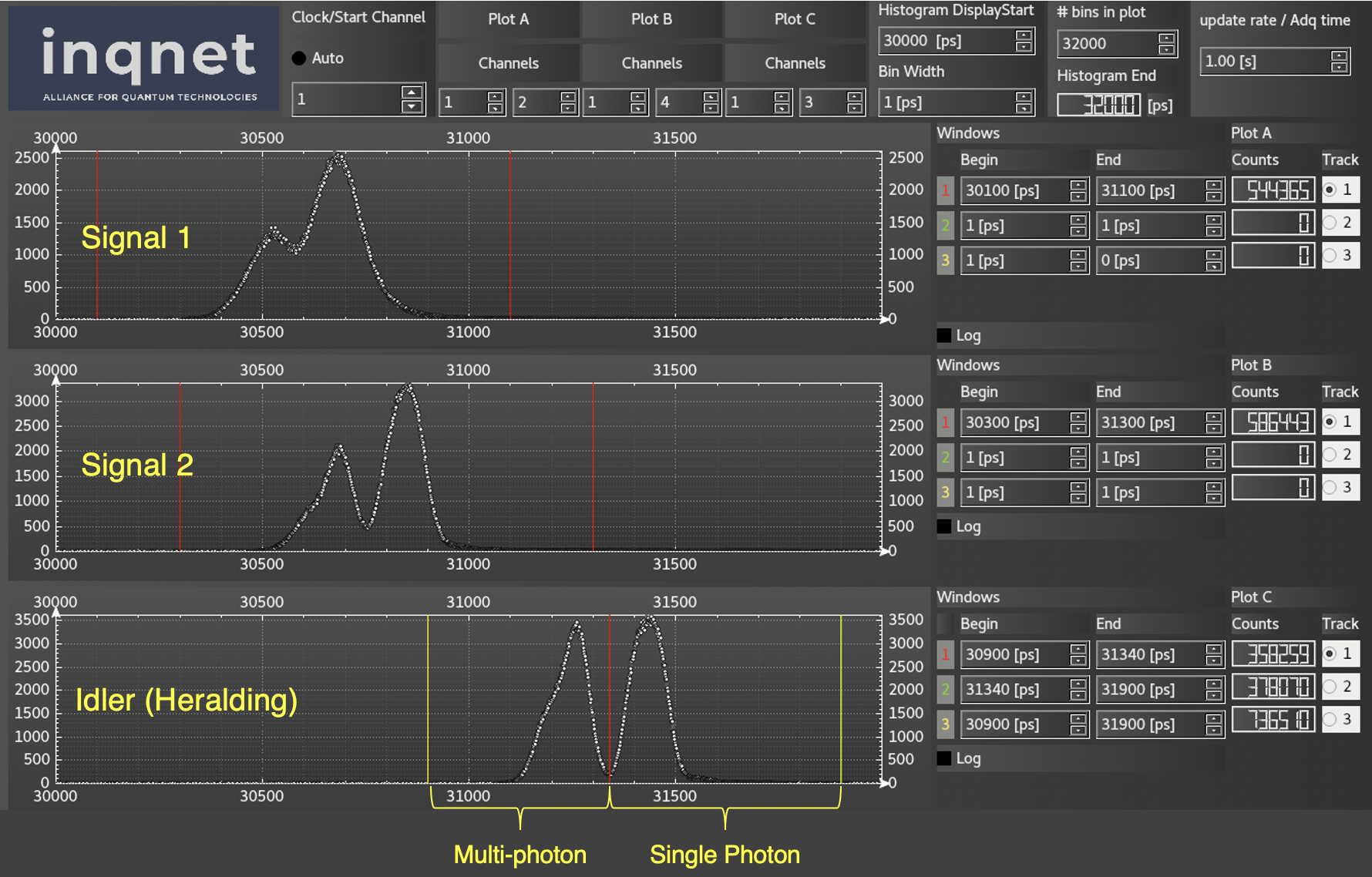}
    \caption{Custom-made Graphical User Interface  allows time-resolved detection of photons and real-time filtering of multi-photon events. The idler mode depicts a bimodal distribution of time tags relative to the clock for an acquisition time of 1 s. The left bin corresponds to the multi-photon events and the right bin corresponds to the single-photon events.}
    \label{fig:gui}
\end{figure}

The readout pulses from the detectors and the clock signal from the AWG are sent to a time tagger that is interfaced with custom-made graphical user interface (GUI) for real-time analysis and multi-photon event discrimination.
The GUI is depicted in Fig. \ref{fig:gui}. 
The recorded detection events in a time bin, that is, the time-tags arriving in a temporal interval defined by the red and yellow markers, are collected over a set acquisition time interval.
A range of potential arrival times of photons in the signal paths are shown in the top two channels of the GUI, and the single and multi-photon events at the idler PNR detector are shown in the bottom channel of the GUI.
The GUI is used to collect single detection events, two-fold coincidence events, and three-fold coincidence events conditioned on the single- and multi-photon detection events at the idler detector.  
In other words, the GUI allows collecting all events for analyzing heralding of photons in the signal path conditioned on threshold and PNR detection of photons in the idler path.

 \subsection{Characterization}
We characterize the setup in two ways: by (1) theoretical calculation and measurement of the joint spectral properties of the photon pair source and (2) by measuring the signal 1, signal 2, and idler path efficiencies from detection rates with low $\mu$, as described below.

\subsubsection{Joint spectral intensity}
\label{sec:photonpairsourcecharacterization}
\begin{figure}[h!]
    \centering
    \includegraphics[width=\columnwidth]{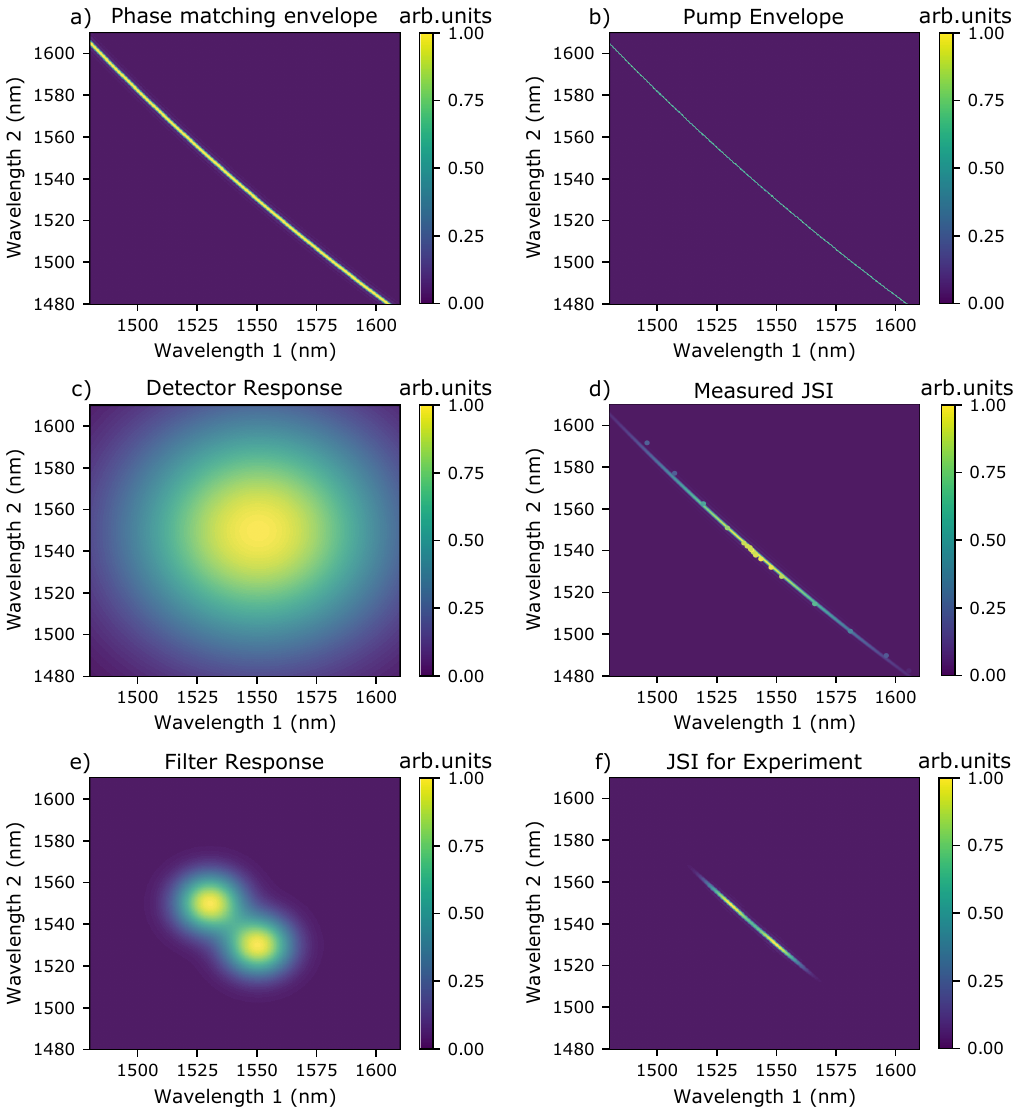}
    \caption{Measured and theoretically calculated joint spectral information used to characterize the photon pair source. a) Phase-matching envelope, b) pump spectral envelope, c) detector response, d) measured (circles) and calculated (contour) JSI, e) filter response of CWDM, and f) JSI for the main experiment, also shown in Fig.~\ref{fig:setup}.
    }
    \label{fig:jsi_sims}
\end{figure}

The two photon component of the quantum state describing SPDC at the output of the PPLN waveguide is
\begin{align*}
    \ket{\Psi} = A\int_{0}^\infty \int_0^\infty f(\omega_1, \omega_2) \hat{a}^\dagger(\omega_1) \hat{a}^\dagger(\omega_2) d\omega_{1}d\omega_{2}\ket{0},
\end{align*}
where $A$ is a constant prefactor that depends on the effective nonlinearity and interaction length, $\hat{a}(\omega_1)$ and $\hat{a}(\omega_2)$ are the signal and idler modes with frequencies $\omega_1$ and $\omega_2$, respectively. 
The joint spectral amplitude (JSA) is
\begin{align*}
    f(\omega_1, \omega_2) = \psi_{ph}(\omega_1, \omega_2)\cdot \psi_p(\omega_1, \omega_2), 
\end{align*}
comprised of the phase-matching and pump envelope amplitudes $\psi_{ph}(\omega_1, \omega_2)$ and $\psi_{p}(\omega_1, \omega_2)$, respectively.
The joint spectral intensity (JSI) is $|f(\omega_1, \omega_2)|^2$. 
We model the phase-matching envelope intensity as
\begin{align*}
    |\psi_{ph}(\omega_1, \omega_2)|^2&=\text{sinc}^2\left( \frac{\Delta k L}{2}\right),
\end{align*}
where $L= 1 \text{ cm}$ is the length of the waveguide and $\Delta k$ is the phase-mismatch.
The calculated phase-matching envelope intensity is depicted in Fig. \ref{fig:jsi_sims}a. 
The phase mismatch for co-linear quasi-phase-matching is 
\begin{align*}
    &\Delta k = 2\pi\left(\frac{  n(\lambda_p)}{\lambda_p}-\frac{ n(\lambda_1)}{\lambda_1} - \frac{n(\lambda_2)}{\lambda_2}-\Gamma\right), 
\end{align*}
where $n_{p(1)(2)}$ is the pump (signal) (idler) index of refraction, $\lambda_{p(1)(2)}=\frac{2\pi c}{\omega_{p(1)(2)}}$ is the pump (signal) (idler) wavelength, $m$ is an integer, $\Lambda$ is the poling period of the crystal, and $\Gamma = m/\Lambda = 400$ mm$^{-1}$ \cite{laudenbach2016modelling}. 
The index of refraction for light of wavelength $\lambda$ in our PPLN waveguide is approximately
\begin{align*}
    n(\lambda)=\sqrt{1+\frac{2.6734 \lambda^{2}}{\lambda^{2}-0.01764}+\frac{1.2290 \lambda^{2}}{\lambda^{2}-0.05914}+\frac{12.614 \lambda^{2}}{\lambda^{2}-474.60}},
\end{align*}
where $n($1540 nm$) = 2.21$ and $n($770 nm$) = 2.26$ \cite{zelmon1997infrared}.
We model the pump envelope intensity as
\begin{align*}
       |\psi_{p}(\omega_1, \omega_2)|^2&= \exp\left(-\frac{(\omega_p -\omega_1- \omega_2)^2}{\sigma_p^2}\right), 
\end{align*}
where $\omega_p = \frac{2\pi c}{770 \text{ nm}}$ and $\sigma_p \sim \frac{2\pi}{100 \text{ ps}}= 60 \text{ GHz}$, as estimated from independent measurements, which is subject to energy conservation $\omega_p = \omega_1 +\omega_2$.
Figure \ref{fig:jsi_sims}b shows the calculated pump envelope intensity.

To characterize the photon pair source, its JSI is determined by performing coincidence measurements of the signal and idler modes after they pass tunable filters with 0.22 nm bandwidths (setup not shown in Fig.~\ref{fig:setup}). 
The measured JSI accounts for the detector response:
\begin{align}
    |f_{m}(\omega_1, \omega_2)|^2 = |\psi_{ph}(\omega_1, \omega_2)|^2\cdot |\psi_p(\omega_1, \omega_2)|^2\cdot |\psi_{d}(\omega_1, \omega_2)|^2, \label{eq:JSI_meas}
\end{align}
where the third factor is the detector efficiency distribution
\begin{align*}
       |\psi_{d}(\omega_1, \omega_2)|^2&=\exp\left({-\frac{( \lambda_1-\lambda_{d})^2 +(\lambda_{2}-\lambda_{d})^2}{\sigma_d^2}}\right), 
\end{align*}
which we model as a Gaussian centered at the optimal detection wavelength of $\lambda_{d}=1550~\text{nm}$ with a spread of $\sigma_d = 53 \text{ nm}$ found from independently performed detector reflectivity measurements.
See Fig. \ref{fig:jsi_sims}c for the calculated detector response.
The measured JSI including detector response is shown in Fig.~\ref{fig:jsi_sims}d using circular markers, with brighter color proportional to the rate of coincidence detection events. 
The contour depicts the theoretical prediction from Eq. \ref{eq:JSI_meas}.

The most relevant JSI is that used for the main experiment, i.e. heralding experiment in configurations (i) with the PNR detector and (ii) with the threshold detector depicted in Fig. \ref{fig:setup}. 
This JSI includes the detector response as well as the response of the CWDM.
The two output modes of the CWDM are centered at $1550$ nm, the idler, and $1530$ nm, the signal, with $\sigma_\text{CWDM}=13$ nm bandwidths. 
Thus, the JSI for the main experiment is modeled as
\begin{align}
    |f_{exp}(\omega_1, \omega_2)|^2 &= |f_{m}(\omega_1, \omega_2)|^2\cdot |\psi_{f}(\omega_1, \omega_2)|^2, \label{eq:JSI_exp}
\end{align}
with the filter response being
\begin{align*}
    |\psi_f(\omega_1, \omega_2)|^2 &
    approx\exp\left(-\frac{(\lambda_1 - \lambda_{\text{f},1})^2+(\lambda_2 - \lambda_{\text{f},2})^2}{\sigma_f^2}\right) 
    \\&+ \exp\left(-\frac{(\lambda_1 - \lambda_{\text{f},2})^2+(\lambda_2 - \lambda_{\text{f},1})^2}{\sigma_f^2}\right),  \nonumber 
\end{align*}
where $\lambda_{\text{f},1} = 1550 \text{ nm}$ and $\lambda_{\text{f},2} = 1530 \text{ nm}$. 
The theoretical response of the CWDM is shown in Fig.~\ref{fig:jsi_sims}e whereas Fig.~\ref{fig:jsi_sims}f depicts the JSI for the main experiment as calculated from Eq.~\ref{eq:JSI_exp}. 

\begin{figure}[h!]
    \centering
    \includegraphics[width=\columnwidth]{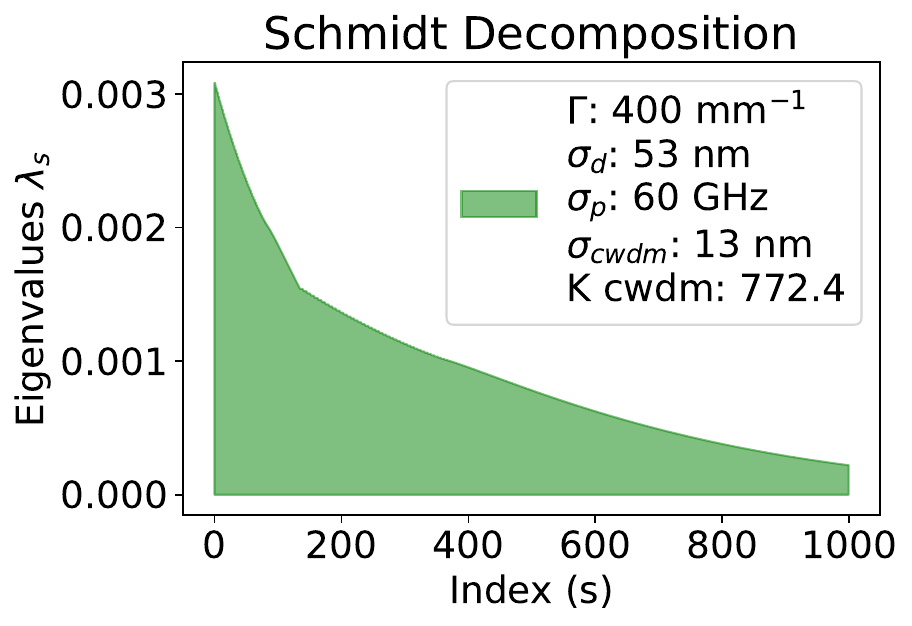}
    \caption{Eigenvalue spectrum $\sum_{s}\lambda_s=1$ obtained from a Schmidt decomposition of the JSI used in the main experiment. 
    }
    \label{fig:schmidt}
\end{figure}

We perform a Schmidt decomposition of the JSI shown in Fig.~\ref{fig:jsi_sims}f by calculating the singular value decomposition of Eq.~\ref{eq:JSI_exp} \cite{zielnicki2018joint}. 
This is relevant for modelling our $g^2(0)$ results, as discussed in Sec.~\ref{sec:model}, and for determining the fidelity of a heralded single photon, see Sec.~\ref{sec:discussion}.
The obtained eigenvalues $\lambda_s$ from the decomposition, normalized by their sum over index $s$, are shown in Fig. \ref{fig:schmidt}, corresponding to a Schmidt number of $K=1/\sum_{s}{\lambda}^2_s \approx 772$.

\begin{figure}[h!]
    \centering
    \includegraphics[width=0.45\columnwidth]{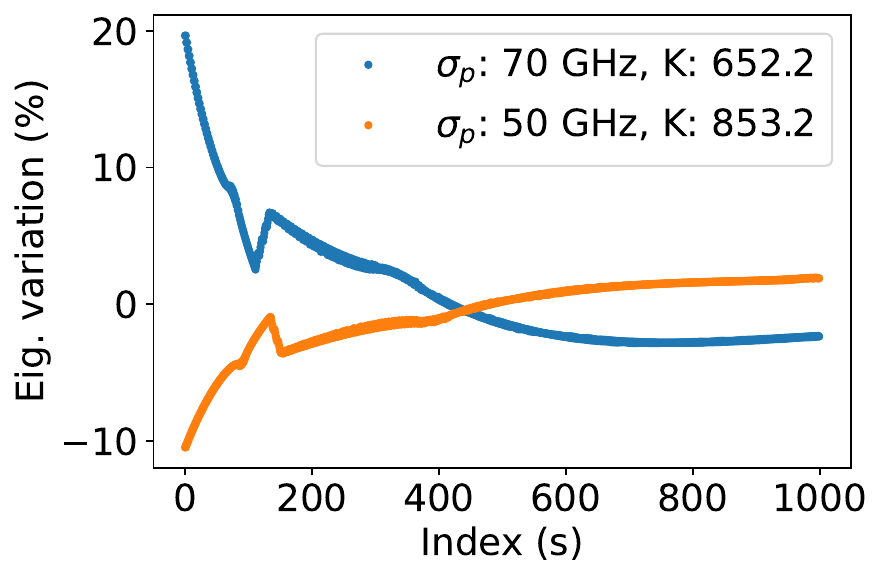}
    \hspace{5pt}
    \includegraphics[width=0.45\columnwidth]{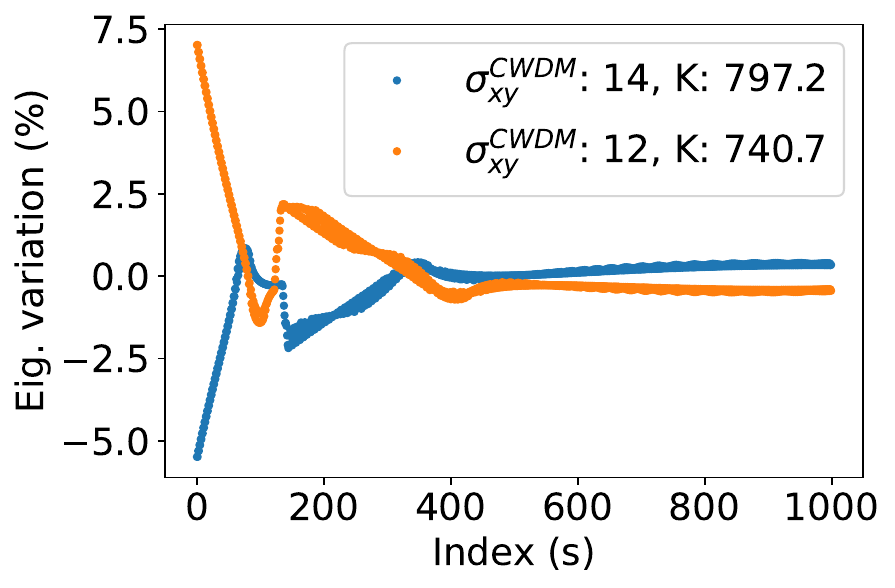}
    \vspace{5pt}
    \includegraphics[width=0.45\columnwidth]{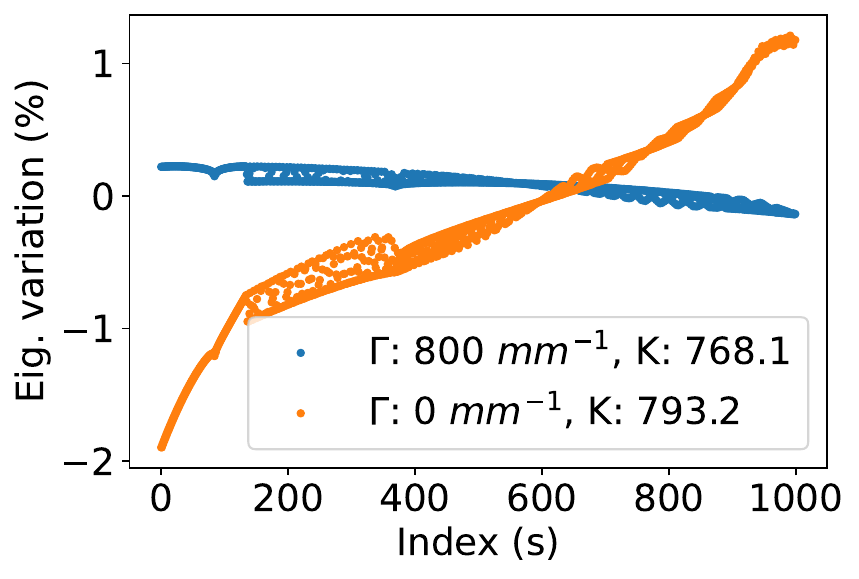}
    \hspace{5pt}
    \includegraphics[width=0.45\columnwidth]{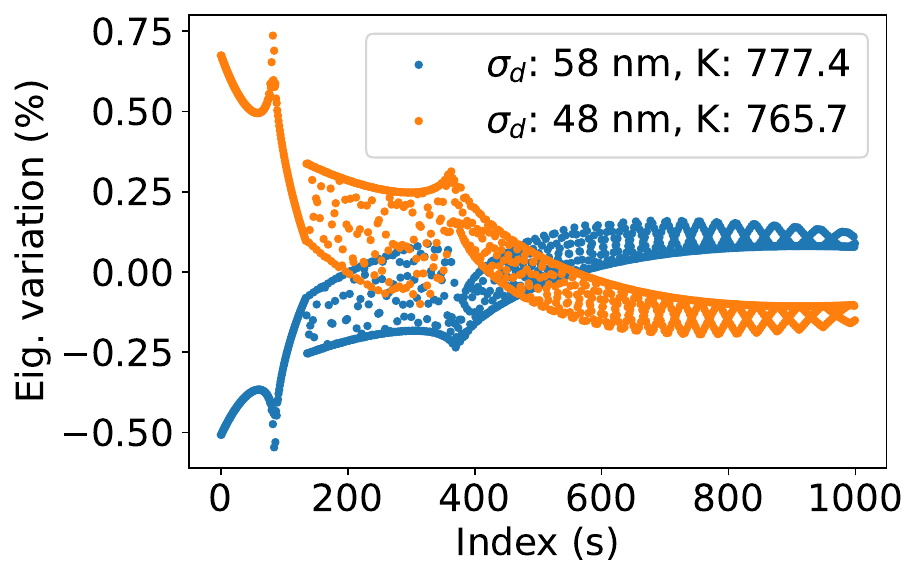}
    \caption{Sensitivity of the Schmidt eigenvalue decomposition of the JSI for the main experiment is probed by varying key parameters of Eq.~\ref{eq:JSI_exp}.
    Central values and variations for each relevant parameter are as follows: $\sigma_p = 60 \pm 10 \text{ GHz}$, $\sigma_\text{CWDM} = 13 \pm 1 \text{ nm}$, $\Gamma = 400 \pm 400 \text{ mm}^{-1}$ and $\sigma_d = 53 \pm 5 \text{ nm}$, with maximum and minimum variations shown.
    These variations are beyond typical experimental uncertainties and are taken as a worst-case scenario.
    The variation of each eigenvalue is normalized to the size of the first eigenvalue $\lambda_0$.
    }
    \label{fig:eig_variation}
\end{figure}

Finally, we determine the sensitivity of the Schmidt decomposition to any potential uncertainty in its key underlying parameters.
We independently vary $\sigma_p$, $\sigma_\text{CWDM}$, $\Gamma$ and $\sigma_d$, see Eq.~\ref{eq:JSI_exp}, and re-calculate Schmidt decomposition, with results shown in Fig.~\ref{fig:eig_variation} and its caption.
We find that, unsurprisingly, the variations of the pump $\sigma_p$ and filter $\sigma_\text{CWDM}$ bandwidths have a significant impact on the Schmidt decomposition \cite{christ2012limits}.
Indeed a single spectral mode can be approximated if $\sigma_p>>\sigma_\text{CWDM}$ \cite{zukowski1995}.
Consequently, the variations of $\sigma_p$ and $\sigma_\text{CWDM}$ have the largest impact on our theoretical model introduced in Sec.~\ref{sec:model}, and are hence propagated in the fit of the model to the data, see Sec.~\ref{sec:results}.

\begin{figure}
    \centering
    \includegraphics[width = \columnwidth]{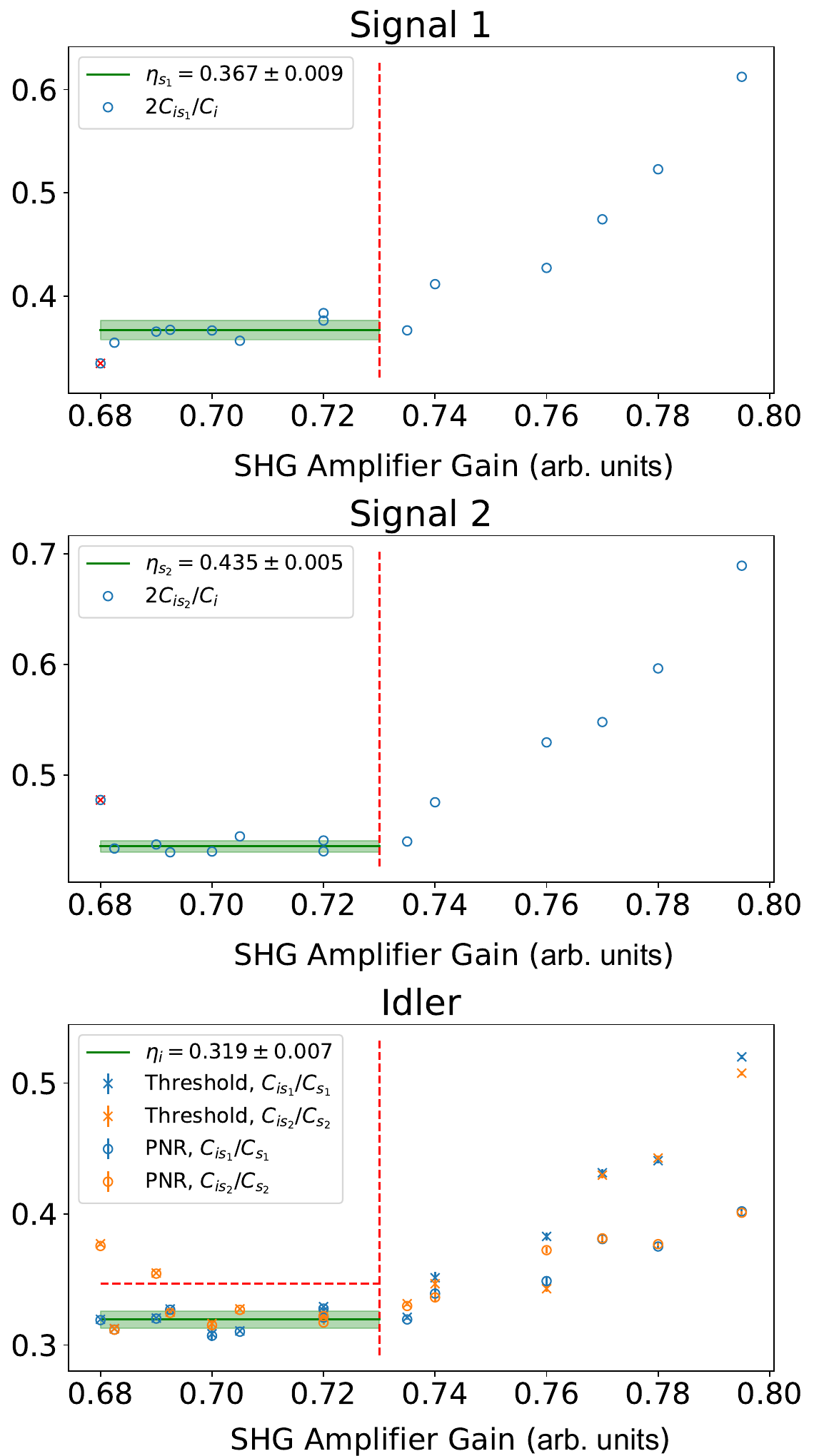}
    \caption{Ratios of single and two-fold coincidence detection rates for the signal and idler paths for varied gain of the amplifier in SHG module.
    The signal 1, signal 2, and idler path efficiencies, are estimated as shown in the insets using the data on the left (and below) of the red dashed lines, which corresponds to $\mu<<1$.
    Idler efficiencies are measured in configurations with the PNR and threshold detector.
    The mean and standard deviation of the fitted efficiencies are indicated by green lines, with numerical values in the insets.
    } 
    \label{fig:eta_estimation}
\end{figure}

\subsubsection{Path efficiencies}
\label{sec:path_efficiencies}
We determine the efficiencies of the signal and idler paths, that is, from PPLN to detection as shown in Fig.~\ref{fig:setup}, by calculating the ratio of coincidence to single photon detection rates using our photon pair source \cite{klyshko1980use}.
The output of our photon-pair source is a two-mode squeezed vacuum (TMSV) state, and can be approximated as
\begin{align*}
    \ket{\psi}_{\text{TMSV}} \approx \sqrt{1-\mu}\ket{0_i0_s} + \sqrt{\mu}\ket{1_i1_s},
\end{align*}
if $\mu<<1$, neglecting loss. 
In this limit, the probability of generating one pair of photons is given by $\mu$, and can be determined by measuring the coincidence-to-accidental ratio~\cite{bussieres2008fast}.
Correspondingly, the relevant detection rates in our main experiment can be approximated to lowest order in $\mu$:
\begin{align}
    C_i &\approx R \eta_i\mu,\label{eq:Ci_low_mu}\\
    C_{s_j} &\approx \frac{1}{2}R\eta_{s_j}\mu,\label{eq:Csj_low_mu}\\
    C_{is_j} & \approx \frac{1}{2}R\eta_{i}\eta_{s_j}\mu\label{eq:Cisj_low_mu},\\
    C_{s_1s_2}&\approx\frac{1}{2}R\eta_{s_1}\eta_{s_2}\mu^2,\\
    C_{is_1s_2}&\approx \frac{1}{2}R\mu^2 \eta_i \eta_{s_1}\eta_{s_2}(2-\eta_i),\label{eq:Cis1s2_low_mu}
\end{align}
where $C_{i}$ is the detection rate of the idler photons, whereas $C_{s_j}$ and $C_{is_j}$ are the detection rates of signal $j$ and idler-signal $j$ two-fold coincidence events, $j=1,2$.
The two-fold coincidence detection rates for photons in the signal 1 and 2 paths is $C_{s_1s_2}$ and $C_{is_1s_2}$ is the rate of three-fold coincidence detection events for photons in the idler and the two signal paths.
The transmission efficiencies of the idler and two signal paths are $\eta_i$ and $\eta_{s_j}$, respectively, and include detector the efficiencies.

To estimate the path efficiencies, we calculate the ratios of two-fold coincidences to single detection rates the signal 1, signal 2, and idler paths, plotting them in Fig.~\ref{fig:eta_estimation} for varied $\mu$.
The amplification in the SHG module is adjusted as a proxy for $\mu$ and a linear fit to the data (green line) is used to obtain the average efficiencies and associated uncertainties at $\mu<<1$, bounded by the red vertical dashed lines.
The mean efficiencies and associated uncertainties (standard deviations) for the signal 1 and 2 paths are $\eta_{s_1}=2C_{is_1}/C_i=0.367\pm 0.009$ and $\eta_{s_2}=2C_{is_2}/C_i=0.435\pm0.005$, respectively.
The idler path efficiency $\eta_i=C_{is_1}/C_{s_1}= 0.319\pm0.007$ is estimated from both the PNR and threshold detector configurations. 
The mean and uncertainty of each path efficiency is used to constraint the fit shown in Fig.~\ref{fig:g2_postfit} of Sec.~\ref{sec:results}.

\section{Theoretical Model}
\label{sec:model}
Photon pair sources from bulk optical nonlinearities are typically operated at $\mu<<1$ to suppress multi-photon events.
The $g^2(0)$ measurement performed in our work extends to large $\mu$, where multi-photon contributions are non-negligible and become significantly suppressed in the (i) PNR detection configuration compared to that using (ii) threshold detection.
To incorporate full multi-photon effects without approximation, we use methods from the phase space formulation of quantum optics to derive an expression for $g^2(0)$ as a function of $\mu$.
We take into account all major imperfections, including coupling and detector inefficiencies. 
We note that our model can be extended to include dark counts, which are negligible for our experiment.

\subsection{Characteristic function-based approach}
\label{sec:overview}
The second order correlation function of photons in the signal 1 and 2 paths conditioned on the detection of photons in the idler path for both detector configurations (i) and (ii) is
 \begin{align*}
    g^2(0) = \frac{C_{is_1s_2}C_i}{C_{is_1}C_{is_2}},
\end{align*}
where $C_{is_1s_2}$ is the rate of three-fold coincidence detection events of photons in the idler and signal 1 and 2 paths, $C_{i}$ is the rate of idler detection events, and $C_{is_j}$ is the two-fold coincidence detection rate of idler and signal $j$ events, where $j=1,2$.
Since we are interested in large $\mu$, we cannot utilize Eqs.~\ref{eq:Ci_low_mu}-\ref{eq:Cis1s2_low_mu}.
Hence, to find an expression for $g^{2}(0)$, we derive analytical expressions for the single path detection rates, two-fold coincidence rates, and three-fold coincidence rates using a characteristic function-based approach \cite{takeoka2015full}. 

A characteristic function for an {$N-$mode} bosonic system is defined as
\begin{align}
    \chi(\xi)=\mathrm{Tr}\left\{\hat\rho\exp( -i(\hat x_1, \hat p_1, \hat x_2, \hat p_2, \dots \hat x_N, \hat p_N)\cdot \xi )\right\},\label{eq:charfunc}
\end{align}
where $\hat\rho$ is the density matrix describing the state of the system, $\hat x_i$ and $\hat p_i$ are the conjugate quadrature operators for mode $i$, 
and $\xi \in \mathbb{R}^{2N}$. 
The quadrature operators can be expressed in term of the bosonic creation and annihilation operators as 
\begin{align*}
    \hat x_i = \frac{1}{\sqrt{2}}\left(\hat a_i^\dagger + \hat a_i \right), \quad \quad  \hat p_i = \frac{i}{\sqrt{2}}\left(\hat a_i^\dagger - \hat a_i \right).
\end{align*}
Equation ~\ref{eq:charfunc} defines a unique mapping from the space of all possible quantum states to a space of functions over $\mathbb{R}^{2N}$, i.e. a quantum system is completely characterized by its characteristic function $\chi(\xi)$ ~\cite{Weedbrook2011}.

A revelant subclass of quantum states is defined by the states whose characteristic function is given by a multivariate Gaussian function:
\begin{align*}
    \chi(\xi)=\exp(-\frac{1}{4}\xi^T\gamma\xi-id^T\xi),
\end{align*}
i.e. they are completely characterized by the displacement vector $d$ and covariance matrix $\gamma$, corresponding to the first and second moments. 
Representatives of this subclass include vacuum, coherent, and thermal states as well as single- and two-mode squeezed states.

Relevant for our experiment is the non-degenerate output of an SPDC process, which can be described as a TMSV state whose covariance matrix is given by
\begin{gather*}
    \gamma_{\mathrm{SPDC}}(\mu) = 
    \begin{pmatrix}
    \mathbf{A}&\mathbf{B}\\
    \mathbf{B}&\mathbf{A}
    \end{pmatrix},\\
    \mathbf{A}=
    \begin{pmatrix}
    1 +2\mu & 0\\
     0 & 1 + 2\mu
    \end{pmatrix},\\
   \mathbf{B}=
    \begin{pmatrix}
    2\sqrt{\mu(\mu+1)} & 0\\
     0 & -2\sqrt{\mu(\mu+1)}
    \end{pmatrix},
\end{gather*}
in block matrix form, where $\mu$ is the mean photon pair number. 
This description is only valid for an SPDC source where only one signal and one idler mode are present. 
If the source allows for multiple signal and idler modes, like in the broadband source we use in our experiment, then the initial state must be modified to include all relevant Schmidt modes, determined through the singular value decomposition of the JSI  \cite{zielnicki2018joint}, as calculated in Sec.~\ref{sec:photonpairsourcecharacterization}.
In this case, the initial state is a product state of the TMSV states in the corresponding Schmidt modes. 
The covariance matrix of the system is then given by a direct sum of the covariance matrices of the respective modes 
\begin{align*}
    \gamma = \gamma_{\mathrm{SPDC}}(\lambda_1\mu)\oplus \gamma_{\mathrm{SPDC}}(\lambda_2\mu) \oplus \dots,
\end{align*}
for an SPDC source that supports $N$ modes with Schmidt coefficients $\lambda_1, \lambda_2,\cdots \lambda_N$, where the sum runs over all relevant modes, $\lambda_1 \geq \lambda_2 \geq \dots \lambda_N$, and $\sum_{s=1}^N \lambda_s = 1$, as before.

Since linear optics preserves the Gaussian nature of states \cite{Weedbrook2011}, i.e. it maps Gaussian states onto Gaussian states, linear optical operations can be described by a symplectic transformation $S$ of the displacement vector and covariance matrix:
\begin{align*}
    &d^\prime = S^Td, &\gamma^\prime = S^T\gamma S.
\end{align*}
For example, the transformation between the input modes $\hat a$, $\hat b$ and the output modes $\hat a^\prime$, $\hat b^\prime$ of a beamsplitter with transmittivity $t$ is given by
\begin{align*}
    \hat a^\prime = t\hat a +i \sqrt{1-t^2}\hat b,\\
    \hat b^\prime = t\hat b + i\sqrt{1-t^2}\hat a.
\end{align*}
We can now find the symplectic transformation $S$ of the beamsplitter that transforms the quadrature operators:
\begin{gather*}
    \begin{pmatrix}
    \mathbf{x}_{a^\prime}\\
    \mathbf{x}_{b^\prime}
    \end{pmatrix}=S^T\begin{pmatrix}
    \mathbf{x}_a\\
    \mathbf{x}_b
    \end{pmatrix}=\begin{pmatrix}
   \mathbf{T}&\mathbf{R}\\
   \mathbf{R}&\mathbf{T}
    \end{pmatrix}\begin{pmatrix}
    \mathbf{x}_a\\
    \mathbf{x}_b
    \end{pmatrix},\\
    \mathbf{x}_i = (\hat{x}_i, \hat{p}_i)^T,\\
    \mathbf{T} = \begin{pmatrix}
    t&0\\
    0&t
    \end{pmatrix}, \quad\mathbf{R} = \begin{pmatrix}
    0&-\sqrt{1-t^2}\\
    \sqrt{1-t^2}&0
    \end{pmatrix},
\end{gather*}
in block matrix form, where $(\mathbf{x}_i, \mathbf{x}_j)^T = (\hat{x}_i, \hat{p}_i, \hat{x}_j, \hat{p}_j)^T$.
The beamsplitter transformation is particularly important because it is used to model path efficiency $\eta_{\mathrm{path}}$, which is reduced from unity by propagation and coupling loss as well as detector inefficiency. 
This is accomplished by combining the mode of interest and vacuum on a beamsplitter of transmittivity $\eta_{\mathrm{ch}}$ and tracing out the reflected mode. 

Given that our setup consists of linear optics, and that loss is modeled as a linear optic transformation, we are able to derive a symplectic transformation $S_{\mathrm{system}}$, with which we calculate the characteristic function of the system up to detection
\begin{align*}
    \gamma_{\mathrm{out}}=S^T_{\mathrm{system}}\gamma_{\mathrm{in}}S_{\mathrm{system}}.
\end{align*}
From the covariance matrix of the final Gaussian state, we can calculate several relevant experimental values such as detection probabilities or rates, which can be used to predict key figures of merit such as fringe visibilities or state fidelities of qubits \cite{valivarthi2020teleportation}. 

Concerning the photon detection step, consider a measurement operator $\hat{\Pi}$.
The probability of detecting the measurement outcome for a given state $\hat{\rho}$ is
\begin{equation}
      Tr[\hat{\rho}\hat{\Pi}] = \bigg(\frac{1}{2\pi}\bigg)^N \int dx^{2N} \chi_{\rho}(x)\chi_{\Pi}(-x),\label{eq:threshold} 
\end{equation}
 where $\chi_{\Pi}(-x)$ is the characteristic function of the measurement operator and is defined in the same way as Eq.~\ref{eq:charfunc} but with $\hat{\Pi}$ instead of $\hat \rho$.
 For threshold detectors, which destructively discriminate between non-zero and zero photons, that is, a detection event and non-event, their measurement operators are
\begin{align*}
      \hat{\Pi}_{\text{no event}}&=\ket{0}\bra{0},\\
     \hat{\Pi}_{\text{event}}&=\hat{I}-\ket{0}\bra{0},
\end{align*}
i.e. we can model the threshold detectors by  projections onto the vacuum state. 
Since the vacuum state is a Gaussian state, the integrand in Eq.~\ref{eq:threshold} is a multi-variant Gaussian function yielding
\begin{align}
    Tr[\hat{\rho}\hat \Pi_{\text{no event}}]=\frac{2^{N_{meas}}}{\sqrt{\text{det}(\gamma_{\mathrm{\text{red}}}+I)}}e^{{-d_{\text{red}}^T (\gamma_{\text{red}}+I_{\text{red}})^{-1}d_{\text{red}}}}, \label{eq:char_int}
\end{align}
where $N_{meas}$ is the number of modes being measured, $\gamma_{\mathrm{red}}$ is the reduced covariance matrix, and $d_{\mathrm{red}}$ is the reduced displacement vector obtained from $\gamma$ and $d$ by tracing out all modes but those measured.

\subsection{Photon-number-resolving detector}
\label{sec:model_PNR_detector}

\begin{figure}[h!]
    \centering
    \includegraphics[width=\columnwidth]{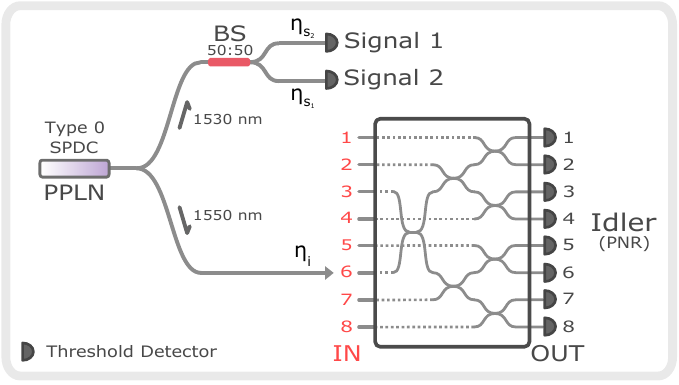}
    \caption{Schematic of the setup used for theoretical modeling. 
    The PNR detector is modeled as a $2N$-port beamsplitter in a binary tree architecture with threshold detectors at the outputs. 
    Only 8 ports are shown in the figure for simplicity.
    The SPDC source and paths depict a simplified representation of that shown in Fig.~\ref{fig:setup}.
    The efficiencies, including all coupling and detection loss, of the signal 1, signal 2, and idler paths, are $\eta_{s_1}$, $\eta_{s_2}$, and $\eta_i$, respectively. 
    }
    \label{fig:model}
\end{figure}

Since the measurement operators describing PNR detectors are not Gaussian operators \cite{leonhardt1997measuring}, we cannot evaluate Eq.~$\ref{eq:char_int}$ to find the probability of detecting one or more photons for the PNR detector.
Instead, we model the PNR detector as an effective $2N$-port beamsplitter with threshold detectors at each output port \cite{fitch2003photon,Paul1996,feito2009measuring,achilles2004photon}. 
We implement the $2N$-port beamsplitter as a network of beamsplitters forming a so-called ``binary tree" architecture, which has $N$ input and output ports, as depicted in Fig. \ref{fig:model} for the case $N=8$.

To model a PNR detector, photons are injected to an input of the ``top-most" beamsplitter of the tree, e.g. input 6 in Fig. \ref{fig:model}.
The detection of photons with the PNR SNSPD is modeled as detection events from any combination of threshold detectors at the output ports of the tree.
For an input Fock state $\ket{n}$, the probability that multiple photons arrive to the same output port is negligible when $N\gg n$, corresponding to ideal photon number discrimination. 
In this case, the number of detection events equals the number of input photons.
For $N\sim n$, the probability of multiple photons arriving to the same output port is non-negligible, corresponding to non-ideal photon number discrimination. 
In this case, the number of detection events does not equal the number of input photons. 
Therefore, the depth of the tree $k=\log_2(N)$, is a figure of merit for photon number discrimination. 

With our beamsplitter tree and threshold detector model, we are able to employ Gaussian characteristic function techniques to find photon detection probabilities, which we later employ to calculate coincidence detection probabilities, and hence detection rates.
We append $N-1$ vacuum modes to the state at idler mode and find the symplectic matrix that transforms the characteristic function of the input state to the tree to that of the output state.
The symplectic matrix $S_\text{k}$ of a beamsplitter tree with depth $k$ can be constructed using the recursive symmetry of the binary tree architecture
\begin{align*}
    S_k = (S_{k-1}\oplus S_{k-1})S_{k,t},
\end{align*}
where $S_{k,t}$ is the symplectic matrix corresponding to the top-most beamsplitter for a tree with depth $k$.
The covariance matrix $\gamma_N$ and displacement vector $d_N$ of the $N$-mode input state to the tree then transform as
\begin{align*}
    &d_N^{\prime} = S_k^Td_N, &\gamma_N^\prime = S_k^T\gamma_N S_k, 
\end{align*}
which is followed by threshold detection at each of the $N$ outputs. 

The probability of detecting non-zero photons at $m$ of the modes, i.e. having a m-fold coincidence event, for an $N$-mode Gaussian state with covariance matrix $\gamma_N$ is
 \begin{align*}
     &\sum_{\{m\}}Tr[\hat{\rho}^{\gamma_N}(\hat{\Pi}_{\text{event}}^{\otimes m}\otimes\hat{\Pi}_{\text{no event}}^{\otimes(N-m)})]\\&\quad\quad\quad=\sum_{\{m\}}Tr[\hat{\rho}^{\gamma_N}((\hat{I}-\ket{0}\bra{0})^{\otimes m}\otimes\ket{0}\bra{0}^{\otimes(N-m)})],\nonumber
 \end{align*}
 where $\sum_{\{m\}}$ indicates a sum over the all possible choices of $m$ output modes. This results in a linear combination of $\binom{N}{m}$ terms of the form
\begin{equation}
    Tr[\hat{\rho}^{\gamma_N}(\hat{I}^{\otimes j}\otimes\ket{0}\bra{0}^{\otimes(N-j)})] = Tr[\hat{\rho}^{\gamma_N}_{\text{red}}\ket{0}\bra{0}^{\otimes(N-j)}],
    \label{eq:partialTraceGamma}
\end{equation}
where $\hat{\rho}_{\text{red}}^{\gamma_N}$ is the  reduced density matrix of the system obtained by taking a partial trace over $j$ modes with $0\leq j \leq N$.
One useful property of Gaussian states is that the covariance matrix of the reduced state after a partial trace is simply the sub-matrix corresponding to the remaining system. 
It can be shown that for an $N$-mode system with covariance matrix $\gamma_N$ and displacement vector $d_N$, the probability of measuring zero photons across the $N$ modes is
  \begin{align*} 
  Tr[\hat{\rho}\hat{\Pi}_{\text{no event}}^{\otimes N}] &= Tr[\hat{\rho}\ket{0}\bra{0}^{\otimes N}]\\
  &=\frac{2^N}{\sqrt{\text{det}(\gamma_{N}+I_N)}}e^{-{d_N^T (\gamma_N +I_N)^{-1}d_N}}.\nonumber
 \end{align*}
 Eq. \ref{eq:partialTraceGamma} then simplifies to
 \begin{align*}
     &Tr[\hat{\rho}^{\gamma_N}_{\text{red}}\ket{0}\bra{0}^{\otimes(N-j)}]=\\  &\quad\quad\quad\frac{2^{N-j}}{\sqrt{\text{det}(\gamma_{N-j}+I_{N-j})}}e^{-{d_{N-j}^T (\gamma_{N-j} +I_{N-j})^{-1}d_{N-j}}},\nonumber
 \end{align*}
 where $I_m$ is the $m$ by $m$ identity matrix, $\gamma_{N-j}$ is the sub-matrix of $\gamma_N$ and $d_{N-j}$ is the sub-vector of $d_N$ corresponding to the remaining subsystem of $N-j$ modes.
 By knowing $\gamma_N$ and $d_N$ of the full $N$-mode system, we can find the m-fold coincidence event probability for arbitrary $m$, where $0\leq m \leq N$.
 
For our experiment, we are interested in the single-photon detection probability $P_1^N$ of the PNR detector.
We model this as the probability that a single output mode of the beamsplitter tree contains a photonic excitation:
 \begin{align}
    P_1^N &=  N Tr[\hat{\rho}^{\gamma_N^\prime}(\hat{\Pi}_{\text{event}}\otimes\hat{\Pi}_{\text{no event}}^{\otimes(N-1)})]\nonumber \\
    &=\frac{N2^{N-1}}{\sqrt{\text{det}(\gamma_{N-1}^\prime+I_{N-1})}}e^{{-{d^\prime_{N-1}}^T (\gamma_{N-1}^\prime +I_{N-1})^{-1}d_{N-1}^\prime}}\nonumber \\
    &\quad- \frac{N2^{N}}{\sqrt{\text{det}({\gamma^\prime_{N}}+I_N)}} e^{{-{d^\prime_{N}}^T (\gamma_{N}^\prime +I_{N})^{-1}d_{N}^\prime}}.\label{eq:P1N}
\end{align}
We can also use the $2N$-port beamsplitter model to describe the photon number discrimination capability of the detector, as discussed in the following sections.

\subsubsection{Photon-number detection}
\label{sec:model_PNR}

When a single-photon is sent to an input of the $2N$-port beamsplitter, the action of the beamsplitter corresponding to unitary $U_N$ splits the photon into an equal superposition of the $N$ output modes. 
An arbitrary Fock state $\ket{n}$ directed to a single input port transforms as
\begin{equation*}
    U_N\ket{n} = \frac{1}{(\sqrt{N})^n}\sum_{j_1+\cdots+j_N = n}\sqrt{\frac{n!}{j_1!\cdots j_N!}} \ket{j_1}\cdots \ket{j_N},
\end{equation*}
Thus, the joint probability of finding $j_1$ photons at output 1, $j_2$ photons at output 2, ..., and $j_N$ photons at output $N$ is
\begin{align*}
    &P_n(j_1,\cdots,j_N) = \frac{1}{N^n}\frac{n!}{j_1!\cdots j_N!}, &\text{where}\quad \sum_{i=1}^N j_i = n.
\end{align*}
The probability that $n$ photons at a single input port trigger an $m$-fold coincidence event, i.e. a detection at each of the $m$ output ports, is
\begin{equation} \label{multinomialPmn}
     P_{m,n}^N = \frac{n!}{N^n}\sum_{j_1+\cdots+j_N = n}^{(m)}\frac{1}{j_1!\cdots j_N!}  = \frac{m!}{N^n}\binom{N}{m} S_2(n,m),
\end{equation}
where $1\leq m \leq n$, the notation $(m)$ refers to the condition that $m$ of $\{j_i\}$ are non-zero, and $S_2(n,m)$ is the Stirling number of the second kind. The Stirling number corresponds to the number of ways of partitioning a set of $n$ elements into $m$ non-empty sets \cite{WolframStirling}. 

As $N\rightarrow \infty$, the $2N$-port model approaches a PNR detector with perfect discrimination efficiency, such that the single-photon detection probability equals the single-photon probability of a general input state,
\begin{equation*}
    \lim_{N\rightarrow \infty}P_1^N = \bra{1}\rho^{\gamma^\prime_N}\ket{1}.
\end{equation*}
For example, from Eq. \ref{eq:P1N}, we can find the probability of a detection event at one output of a tree with depth $k$ for an input thermal state with mean photon number $\mu$ as
\begin{align*}
    P_{1}^{k}=\frac{2^{k} \mu}{(1+\mu)\left(2^{k}+\left(2^{k}-1\right) \mu\right)}.
\end{align*}
Similarly, for a coherent state with mean photon number $|\alpha|^2$ as
\begin{align*}
    P(1)_{k}=2^{k} e^{-|\alpha|^{2}}\left(e^{|\alpha|^{2} / 2^{k}}-1\right).
\end{align*}
By taking the limit $k\rightarrow\infty$, we recover the single photon probabilities for a thermal state and coherent state, respectively, as
\begin{align*}
    &\lim _{k \rightarrow \infty} \frac{2^{k} \mu}{(1+\mu)\left(2^{k}+\left(2^{k}-1\right) \mu\right)}=\frac{\mu}{(1+\mu)^{2}},\\
    &\lim _{k \rightarrow \infty} 2^{k} e^{-|\alpha|^{2}}\left(e^{|\alpha|^{2} / 2^{k}}-1\right)=e^{-|\alpha|^{2}}|\alpha|^{2}.
\end{align*}

\subsubsection{POVM elements and counting statistics}
\label{sec:POVM_stats}
The values of $P_{m,n}^N$, from Eq.~\ref{multinomialPmn}, correspond to the matrix elements of a conditional probability matrix $\mathbf{C}$, following the definition used in Ref.~\cite{achilles2004photon}. 
The rows correspond to the positive-operator value measure (POVM) elements of the measurement outcomes and the columns correspond to the Fock projection operators. 
 The matrix for a threshold detector, in other words, a tree with $k=0$ is
\[
\begin{blockarray}{cccccccccccc}
&&\ket{0}\bra{0}&\ket{1}\bra{1}&\ket{2}\bra{2}&\ket{3}\bra{3}&\ket{4}\bra{4}&\ket{5}\bra{5}&\ket{6}\bra{6}&\cdots \\
\begin{block}{cc(cccccccccc)}
   \hat{\Pi}_{\text{no event}} && 1 & 0 & 0 & 0 & 0 & 0 & 0  & \cdots& \\
   \hat{\Pi}_{\text{event}} && 0 & 1 & 1 & 1 & 1 & 1 & 1 & \cdots&\\
\end{block}
\end{blockarray},
\]
with measurement outcomes (rows) and projections (columns) indicated.
The matrix for an ideal PNR detector is the identity matrix
\[
\begin{blockarray}{cccccccccccc}
&&\ket{0}\bra{0}&\ket{1}\bra{1}&\ket{2}\bra{2}&\ket{3}\bra{3}&\ket{4}\bra{4}&\ket{5}\bra{5}&\ket{6}\bra{6}&\cdots& \\
\begin{block}{cc(cccccccccc)}
   \hat{\Pi}_{\text{0}} && 1 & 0 & 0 & 0 & 0 & 0 & 0 &  \cdots& \\
   \hat{\Pi}_{\text{1}} && 0 & 1 & 0 & 0 & 0 & 0 & 0 & \cdots&\\
   \hat{\Pi}_{\text{2}} && 0 & 0 & 1 & 0 & 0 & 0 & 0 &  \cdots&\\
   \hat{\Pi}_{\text{3}} && 0 & 0 & 0 & 1 & 0 & 0 & 0 & \cdots&\\
   \hat{\Pi}_{\text{4}} && 0 & 0 & 0 & 0 & 1 &  0 & 0 &  \cdots&\\
   \hat{\Pi}_{\text{5}} && 0 & 0 & 0 & 0 & 0 & 1 & 0 &  \cdots&\\
   \hat{\Pi}_{\text{6}} && 0 & 0 & 0 & 0 & 0 & 0 & 1 &  \cdots&\\
   \vdots && \vdots & \vdots & \vdots & \vdots & \vdots & \vdots &\vdots &   \ddots&\\
\end{block}
\end{blockarray},
 \]
 For a detector with efficiency $\eta_{d}$, i.e. modelled as path loss of transmittivity $\eta_{d}$ before an ideal detector, the probability that $n$ photons trigger an $m$-fold coincidence detection event is
\begin{align*}
   P_{m,n}^N(\eta_{d})=
   &\sum_{j=0}^n P_{m,j}^N \binom{n}{j} {\eta_{d}}^j(1-{\eta_{d}})^{n-j}\nonumber \\
   &=\sum_{j=0}^n C_{m,j} L_{j,n} \nonumber \\
   &= \left(\mathbf{C}\cdot \mathbf{L}\right)_{m,n},
\end{align*}
where $\mathbf{L}$ is the loss matrix with matrix elements,
\begin{align*}
    L_{j,n} = \binom{n}{j} {\eta_{d}}^j(1-\eta_{d})^{n-j}.
\end{align*}
The matrix corresponding to $\mathbf{C}\cdot\mathbf{L}$ for a tree with $k=3$ and $\eta_{d} = 0.71$ is
\[
\begin{blockarray}{cccccccccccc}
&&\ket{0}\bra{0}&\ket{1}\bra{1}&\ket{2}\bra{2}&\ket{3}\bra{3}&\ket{4}\bra{4}&\ket{5}\bra{5}&\ket{6}\bra{6}&\cdots& \\
\begin{block}{cc(cccccccccc)}
   \hat{\Pi}_{\text{0}} &&  1  & 0.29 & 0.084 & 0.024 & 0.007 & 0.002 & 0.001 &\cdots& \\
   \hat{\Pi}_{\text{1}} && 0  & 0.71 & 0.475& 0.240 & 0.108& 0.046 & 0.019 &  \cdots&\\
   \hat{\Pi}_{\text{2}} &&  0  & 0  & 0.441 & 0.501 & 0.383 & 0.246 & 0.144& \cdots&\\
   \hat{\Pi}_{\text{3}} && 0  & 0 & 0  & 0.235& 0.398 & 0.425 & 0.368 & \cdots&\\
   \hat{\Pi}_{\text{4}} && 0  & 0  & 0  & 0  & 0.104& 0.244 & 0.346 & \cdots&\\
   \hat{\Pi}_{\text{5}} &&  0  & 0  & 0  & 0  & 0  & 0.037 & 0.114 &\cdots&\\
   \hat{\Pi}_{\text{6}} && 0  & 0  & 0  & 0  & 0  & 0  & 0.010& \cdots&\\
   \vdots& & \vdots & \vdots & \vdots & \vdots & \vdots & \vdots & \vdots & \ddots&\\
\end{block}
\end{blockarray}\label{eq:exp_cond_mat},
 \]
 As we will show in detail in Sec.~\ref{sec:results}, this matrix corresponds to our experimental PNR configuration.
 
 The counting statistics $p(n)$ can be related to the input photon number distribution $\varrho(n)$ by
\begin{align*}
    p_m = \sum_n \sum_{j=0}^n C_{m,j} L_{j,n} \varrho_n,
\end{align*}
where $p_m = p(m)$ and $\varrho_n = \varrho(n)$, following the notation of Eq.~9 from Ref.~\cite{achilles2004photon}.
In matrix notation this is
    $\vec{p} = \mathbf{C}\cdot \mathbf{L}  \vec{\varrho}$.
The transpose of the matrix $(\mathbf{C}\cdot \mathbf{L})^T$ is matrix $\mathbf{B}$ from Ref. \cite{feito2009measuring}, which relates probabilities and density matrices as
    $\vec{p} = \mathbf{B} \hat{\rho}$.

\subsubsection{Photon-number discrimination efficiency}
\label{sec:pnr_metric}
A key figure of merit of our detector PNR configuration is its ability to discriminate single photon events from others.
To quantify this, we define the ``$m-$photon discrimination efficiency"  and use it to calculate the ``single-photon discrimination efficiency"  as follows.

A POVM element corresponding to the $m$-photon outcome for a  non-ideal PNR detector can be described by
\begin{align}
    \hat{\Pi}_m &=\sum_{n=0}^\infty c^m_n\ket{n}\bra{n}, \label{eq:POVM_nonideal}
\end{align}
where $c^m_n$ are the matrix elements corresponding to the representation of the operator in the photon number basis, and are each equal to the probability of registering $m$ photons given $n$ incident photons. 
The $m$-photon outcome for an ideal PNR detector is
\begin{align}
    \hat{\Pi}^{ideal}_m=\ket{m}\bra{m}. \label{eq:POVM_ideal}
\end{align}
Note that for a threshold detector, the $m$-photon outcome for $m>0$  is the ``event" outcome $    \hat{\Pi}_{\text{event}} = \sum_{n=1}^{\infty} \ket{n}\bra{n}$, and $m=0$ outcome corresponds to ``no-event" $\hat{\Pi}_{\text{no event}} = \ket{0}\bra{0}$.
We define the $m$-photon discrimination efficiency as \begin{align}
    \eta_{PNR}^m=1-\frac{1}{2} \operatorname{Tr}\left[\sqrt{\left(\frac{\hat{\Pi}_{m}}{\text{Tr}[\hat{\Pi}_{m}]}-\frac{\hat{\Pi}_{m}^{\text {ideal }}}{\text{Tr}[\hat{\Pi}_{m}^{\text {ideal }}]}\right)^{2}}\right], \label{eq:eta_pnr}
\end{align}
where the second term is the trace distance between elements $\hat{\Pi}_m$ and $\hat{\Pi}_m^{ideal}$, normalized by their trace, corresponding to the $m$-photon measurement outcome of the PNR detector.
Using Eqs. \ref{eq:POVM_nonideal} and \ref{eq:POVM_ideal}, we simplify Eq.~\ref{eq:eta_pnr} to 
\begin{align}
   \eta_{PNR}^m =\frac{c^m_m}{\sum_{n=0}^{\infty} c^m_n}=\frac{P(m|m)}{\sum_{n=0}^\infty P(m|n)}, \label{eq:eta_pnr_simp}
\end{align}
where $c_{n}^m = P(m|n)$ is the probability that the detector registers $m$ photons given that $n$ photons were incident on the detector. 
Relevant to our experiment is the single photon discrimination efficiency ($m=1$). As defined in Eq. \ref{eq:eta_pnr_simp}, $\eta_{PNR}^1$ is zero for a threshold detector and unity for an ideal PNR detector. 

\subsection{Analytical expressions of detection probabilities}
\label{sec:analytical_expressions}
For a 2N-port beamsplitter realized as a finite-depth binary tree, we derive the following expressions for detection probabilities of the signal and idler paths, as well as two-fold and three-fold coincidence event probabilities as a function of the efficiencies and tree depth $k$, where $N=2^k$. 
The equations reduce to the threshold detection case for $k=0$.

We use $\hat{\Pi}_{\text{no event},m}$ and $\hat{\Pi}_{\text{event},m}$ to denote the measurement operators for a threshold detector at the $m$th tree output:
\begin{align*}
    \hat{\Pi}_{\text{no event},m} &= \ket{0}\bra{0}_m,\\
    \hat{\Pi}_{\text{event},m} &= \hat{I}_m-\ket{0}\bra{0}_m.
\end{align*}
For the PNR detector, we use $\hat{\Pi}_{\text{event},m}\otimes \hat{\Pi}_{\text{no event}}^{\otimes N-1}$ to denote an ``event" measurement outcome for a detector at the $m$th output and ``no-event" measurement outcomes for the detectors at the remaining $N-1$ outputs of the tree.

\subsubsection{Detection probabilities for signal and idler detectors}
The probabilities $P_{s_1}$ and $P_{s_2}$ of a detection event for the signal 1 and 2 detectors, respectively, are
\begin{align*}
    P_{s_1} &= \text{Tr}\left[\rho\left(\hat{\Pi}_{s_1, \text{event}}\otimes \hat{I}_{s_2}\otimes \hat{I}^{\otimes N}\right)\right],\\
    P_{s_2} &= \text{Tr}\left[\rho\left( \hat{I}_{s_1}\otimes\hat{\Pi}_{s_2, \text{event}}\otimes \hat{I}^{\otimes N}\right)\right],
\end{align*}
and evaluate to
\begin{align*}
    P_{s_j}=1-\prod_s\frac{2}{2+\eta_{s_j}\lambda_s\mu}, 
\end{align*}
where $j=1,2$ and $\lambda_s$ are the Schmidt coefficients obtained from the singular value decomposition of the JSI as discussed in Sec.~\ref{sec:photonpairsourcecharacterization}.  
The products in the expressions run over all Schmidt coefficients.
The mean number of pairs $\mu$ as well as path efficiencies $\eta_{i}$ and $\eta_{s_j}$, where $j=1,2$, as depicted in Fig.~\ref{fig:model} and used here and in the following, are as defined earlier.

The probablity $P_i$ of a detection event for the idler detector is then
\begin{align*}
    P_i
    &=  N  \text{Tr}\left[\rho \left(I_{s_1}\otimes I_{s_2}\otimes \hat{\Pi}_{\text{event},m}\otimes \hat{\Pi}_{\text{no event}}^{\otimes N-1}\right)\right],
\end{align*}
and evaluates to
\begin{align}
    P_i&= 2^k\left(\prod_{s}\frac{2^k}{2^k + (2^k - 1)\lambda_s\mu\eta_i} -\prod_{s} \frac{1}{1+\lambda_s \mu \eta_i}\right). \label{eq:idler_pnr_prob}
\end{align}

\begin{widetext}
\subsubsection{Two-fold coincidence detection probabilities}
The probabilities of a two-fold coincidence detection event at the idler and one of the signal detectors, $P_{is_1}$ and $P_{is_2}$, are
\begin{align*}
    P_{is_1}
    &=  N  \text{Tr}\left[\rho\left( \hat{\Pi}_{\text{event},s_1}\otimes I_{s_2}\otimes \hat{\Pi}_{\text{event},m}\otimes \hat{\Pi}_{\text{no event}}^{\otimes N-1}\right)\right],\\
    P_{is_2}
    &=N  \text{Tr}\left[\rho \left( I_{s_1}\otimes \hat{\Pi}_{\text{event},s_2}\otimes \hat{\Pi}_{\text{event},m}\otimes \hat{\Pi}_{\text{no event}}^{\otimes N-1}\right)\right], 
\end{align*}
and evaluate to
\begin{align}
    P_{is_j}&= 2^k\bigg(\prod_s \frac{2^k}{2^k + (2^k -1)\lambda_s\mu\eta_i}- \prod_s\frac{2^{k+1}}{\lambda_s \mu \eta_{s_j}(2^k - (2^k - 1)\eta_i)+ 2(2^k + (2^k -1)\lambda_s \mu \eta_i)}\nonumber\\
    &-\prod_s\frac{1}{1+\lambda_s\mu\eta_i}+\prod_s\frac{2}{2+2\lambda_s\mu\eta_i + \eta_{s_j}\lambda_s\mu(1-\eta_i)}\bigg),
\label{eq:twofold_coin_prob} 
\end{align}
where $j=1,2$.

The probability of a two-fold coincidence detection event at the signal 1 and 2 detectors is
\begin{align*}
    P_{s_1 s_2}
    &=  \text{Tr}\left[\rho \left(\hat{\Pi}_{\text{event},s_1}\otimes \hat{\Pi}_{\text{event},s_2}\otimes \hat{I}_m^{\otimes N}\right)\right],
\end{align*}
and evaluates to
\begin{align*}
    P_{s_1s_2}&=1-\prod_s\frac{2}{2+\eta_{s_1} \lambda_s\mu}-\prod_s\frac{2}{2+\eta_{s_2} \lambda_s\mu}+\prod_s\frac{2}{2+(\eta_{s_1} +\eta_{s_2} )\lambda_s\mu}.
\end{align*}

\subsubsection{Three-fold coincidence detection probabilities}
The probability of a three-fold coincidence detection event at the idler, signal 1, and signal 2 detectors, $P_{i,s_1,s_2}(\mu, \eta_{s_1}, \eta_{s_2},\eta_{i},k)$, is  
\begin{equation*}
    P_{i,s_1,s_2} = N \text{Tr}\left[\rho\left( \hat{\Pi}_{\text{event}, s_1}\otimes \hat{\Pi}_{\text{event}, s_1}\otimes \hat{\Pi}_{\text{event},m}\otimes \hat{\Pi}_{\text{no event}}^{\otimes N-1}\right)\right],
\end{equation*}
and evaluates to
\begin{align}
      &P_{i,s_1,s_2}=2^k\bigg(\prod_s\frac{2^{k}}{(2^k + (2^k -1)\lambda_s\mu\eta_i)}-\prod_s\frac{2^{k+1}}{\lambda_s\mu\eta_{s_1}(2^k - (2^k - 1)\eta_i)+2(2^k + (2^k -1)\lambda_s\mu\eta_i)}\nonumber \\
      &- \prod_s\frac{2^{k+1}}{\lambda_s\mu\eta_{s_2}(2^k - (2^k - 1)\eta_i)+2(2^k + (2^k -1)\lambda_s\mu\eta_i)}\nonumber+\prod_s \frac{2^{k+1}}{\lambda_s\mu(\eta_{s_1}+\eta_{s_2})(2^k - (2^k - 1)\eta_i)+2(2^k +(2^k - 1)\lambda_s \mu \eta_i)}\nonumber \\
    &-\prod_s\frac{1}{1+\lambda_s\mu \eta_i}\nonumber+\prod_s\frac{2}{2+2\lambda_s\mu\eta_i + \eta_{s_1}\lambda_s\mu(1-\eta_i)}\nonumber+\prod_s\frac{2}{2+2\lambda_s\mu\eta_i + \eta_{s_2}\lambda_s\mu(1-\eta_i)}\\&
    -\prod_s\frac{2}{2+2\mu\lambda_s\eta_i + (\eta_{s_1}+\eta_{s_2})\mu\lambda_s(1-\eta_i)}\bigg).\label{eq:threefold_prob}
\end{align}

\subsubsection{Second-order correlation function $\mathit{g^{(2)}(0)}$}
Finally, we readily derive the analytical expression for $g^{(2)}(0)$ by substituting Eqs.~\ref{eq:idler_pnr_prob}, \ref{eq:twofold_coin_prob}, and \ref{eq:threefold_prob} and into
\begin{align}
    g^{(2)}(0) = \frac{P_{i,s_1,s_2}P_i}{P_{i,s_1}P_{i,s_2}} =\frac{C_{i,s_1,s_2}C_i}{C_{i,s_1}C_{i,s_2}},\label{eq:g_20_analytical}
\end{align}
where the respective probabilities $P$ can be used to calculate detection rates $C$ using $C=R P$.
\end{widetext}

\section{Results}
\label{sec:results}
\begin{figure}[h!]
    \centering
    \includegraphics[width=\columnwidth]{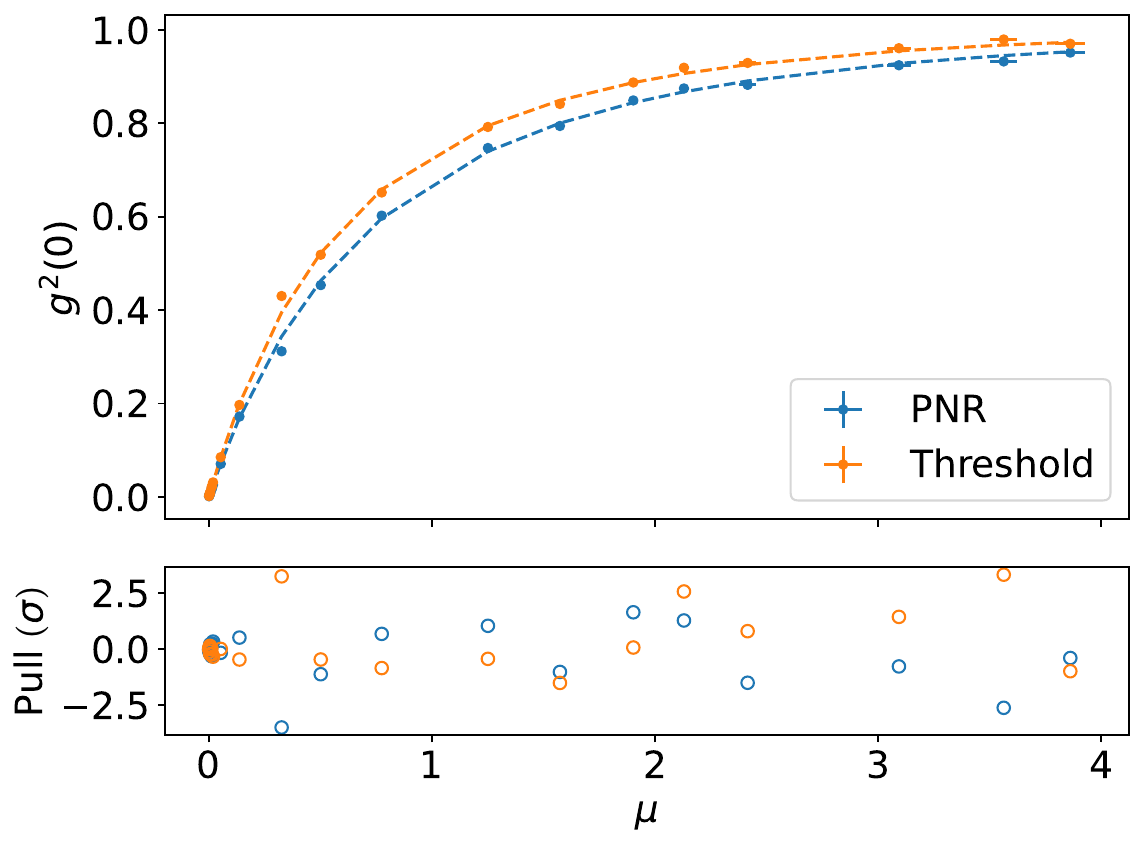}\llap{\raisebox{2.65cm}{\includegraphics[width=0.18\columnwidth]{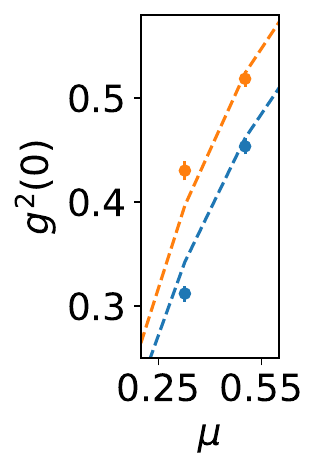}\hspace*{2.88cm}}}
    \caption{Measured correlation function $g^2(0)$ as a function of mean photon-pair number $\mu$.
    The experimental data using threshold (orange) and PNR (blue) idler detector configurations are represented by the circular markers whereas the respective fitted models are shown by dashed lines. 
    Uncertainties of $g^2(0)$, referred to as $\sigma_{g^2(0)}$, are derived from the statistical uncertainties of the coincidence detection events whereas the uncertainties of $\mu$, that is, $\sigma_{\mu}$, are extracted from the fit to the model.
    The inset depicts the region where the largest reduction in $g^2(0)$ is measured.
    The pull between the measured $g^2(0)$ and the model, computed as $[\sigma_{g^2(0)}^2 + \left|\partial_\mu{g^2(0)}\right|^2\sigma_{\mu}^2]^{1/2}$ and plotted in units of standard deviations, is shown at the bottom of the canvas.
    }
    \label{fig:g2_postfit} 
\end{figure}

We vary the gain of the amplifier in the SHG module and measure single detector, i.e. signal 1 and 2 and idler, events as well as two- and three-fold coincidence detection events for (i) the PNR configuration and (ii) the threshold configuration for the idler detector.
We then perform a maximum-likelihood fit of our theoretical model for $g^2(0)$, i.e. Eq.~\ref{eq:g_20_analytical}, to the measured $g^2(0)$ for configurations (i) with PNR and (ii) with threshold detection.
The likelihood is optimized using the MINUIT~\cite{James:1975dr} implementation in \textit{iminuit}~\cite{iminuit}.
The experimental $g^2(0)$ data and curve corresponding to the best-fitted model are shown in Fig.~\ref{fig:g2_postfit}. 

\subsection{Maximum-likelihood fit}
The theoretical model for $g^2(0)$ includes several parameters, as introduced in Secs.~\ref{sec:experimentalmethods} and \ref{sec:model}.
Our fit extracts the following key experimental values: 
mean photon-pair number $\mu$, tree depth $k$, path efficiencies $\eta_i$, $\eta_{s_1}$ and $\eta_{s_2}$, as well as the filter and pump bandwidths, $\sigma_\text{CWDM}$ and $\sigma_p$, which strongly influence the eigenvalue spectrum of JSI.
The best-fit values and uncertainties of the mean photon number for each amplifier setting is shown in Fig.~\ref{fig:g2_postfit}.
We also calculate the pull for $g^2(0)$, which is the distance of the best fit value from the Gaussian constraint measured in values of the constraint width.
The best-fit, uncertainties and pull of the other values are shown in Tab.~\ref{tab:fit_results}.
We find the best-fit path efficiencies and the filter bandwidth are identical, within uncertainty, to that evaluated by independent measurements in Secs.~\ref{sec:path_efficiencies} and \ref{sec:photonpairsourcecharacterization}.
The predicted pump bandwidth (88 GHz) is larger than that measured in Sec.~\ref{sec:photonpairsourcecharacterization} (60 GHz) likely because it was inferred by measurements at telecommunication wavelength.

\begin{figure}[h!]
    \centering
    \includegraphics[width = \columnwidth]{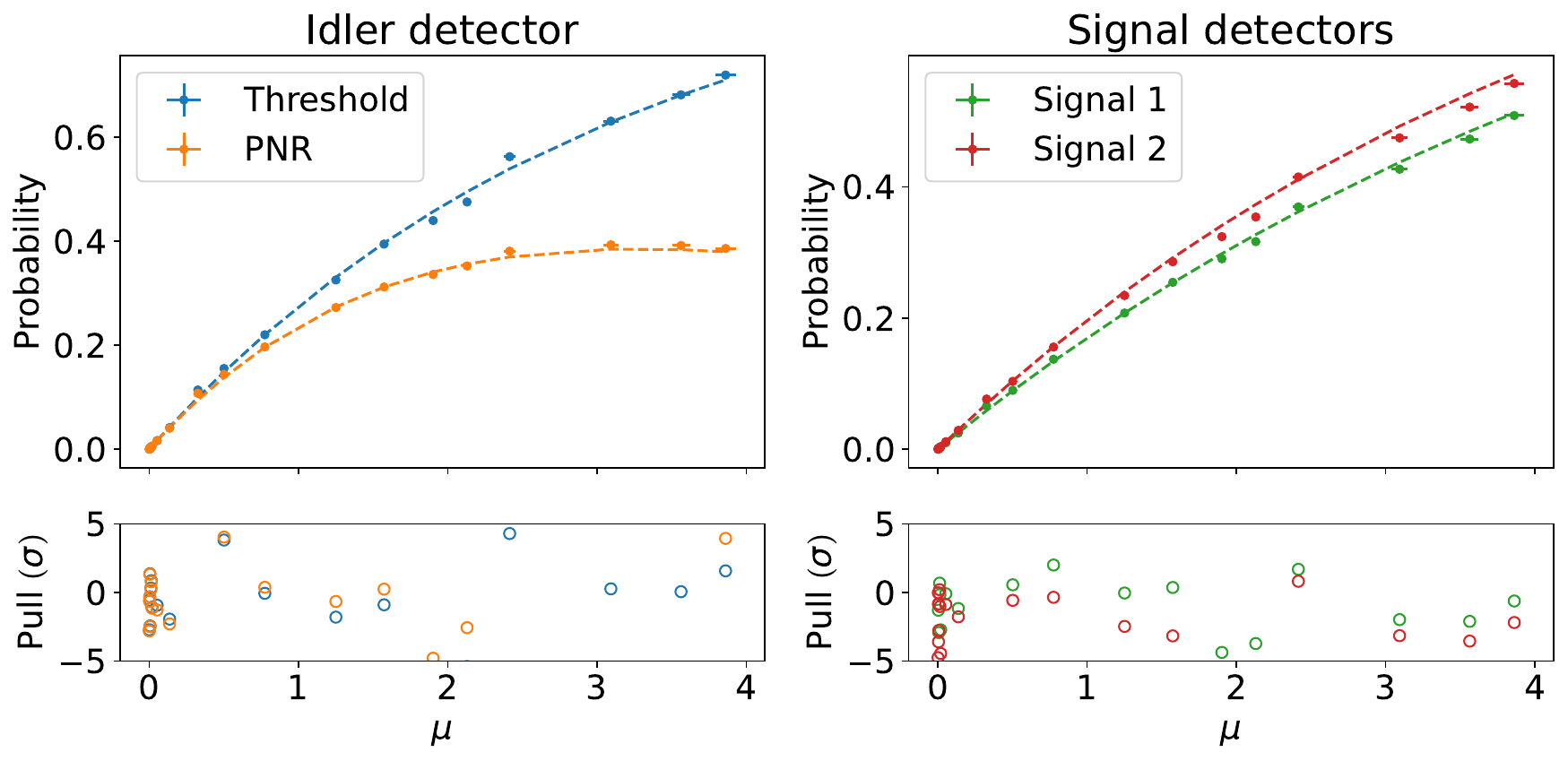}
    \vspace{10pt}
    \includegraphics[width = \columnwidth]{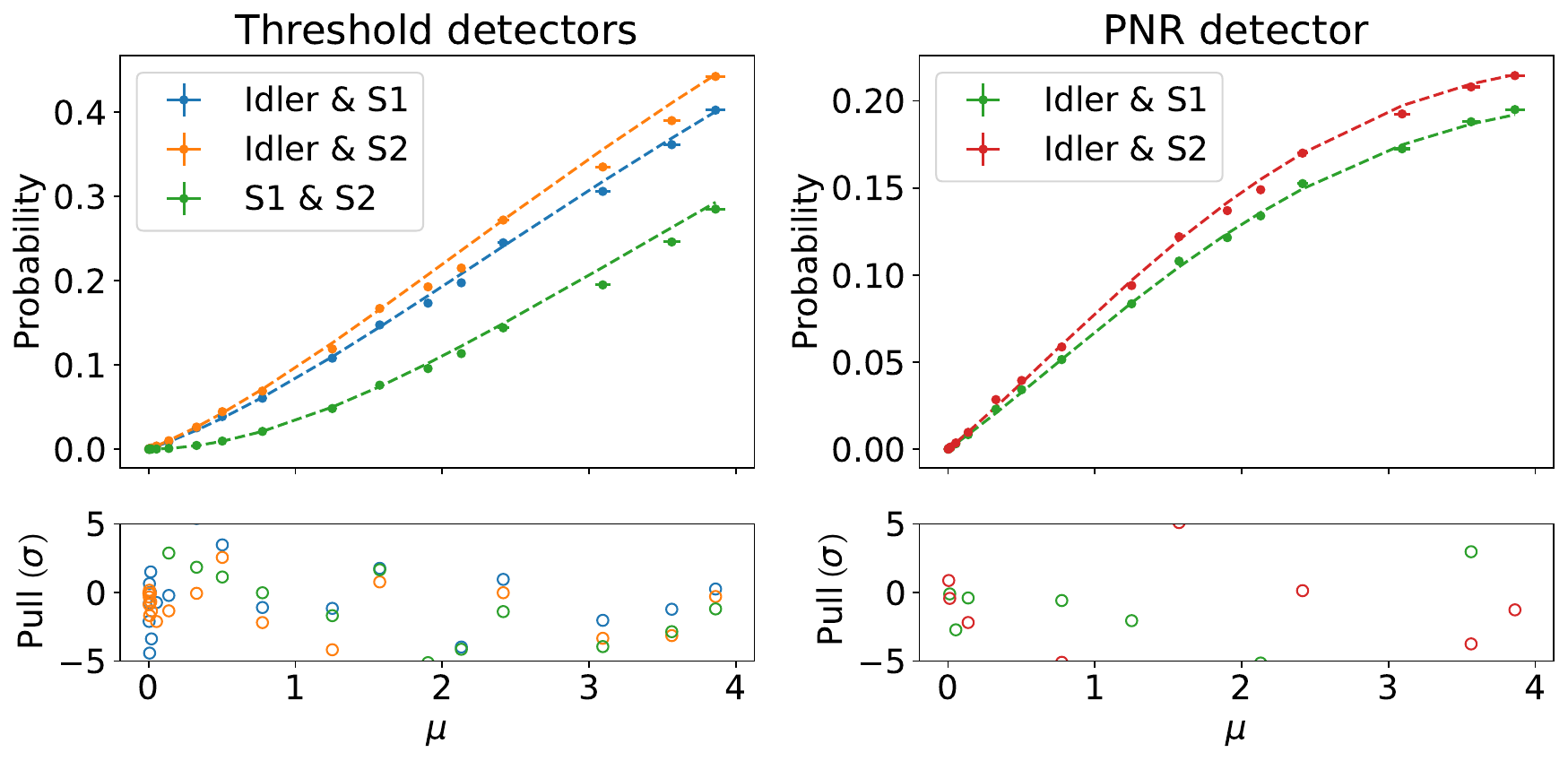}
    \caption{{
    Idler detection probabilities (top left), signal 1 and signal 2 detection probabilities (top right), signal detector two-fold coincidence probabilities and signal detectors with idler in threshold configuration (bottom left), two-fold signal and idler in PNR configuration two-fold coincidence probabilities. The prediction from the theoretical model is shown for the best fit parameters in Tab.~\ref{tab:fit_results}.
    }} 
    \label{fig:rates}
\end{figure}

In the fit, the path efficiencies are free parameters, while the mean and uncertainties thereof, measured in Sec.~\ref{sec:path_efficiencies}, are used to place Gaussian constraints on the fit. 
Each measured $g^2(0)$ is ascribed an independent value of $\mu$, and given the path efficiencies, is determined by fitting the single detector and two-fold coincidence detection probabilities, i.e. those shown in Fig.~\ref{fig:rates}, collected during the measurements.
The mean and statistical uncertainties of these detection rates is used to place a Gaussian constraint on the value of $\mu$ for each data point.
The eigenvalue spectrum of the JSI is computed by varying $\sigma_\text{CWDM}=13 \pm 1 \text{ nm}$ and $\sigma_p= 60 \pm 10 \text{ GHz}$ as discussed in Sec.~\ref{sec:photonpairsourcecharacterization}, and a linear approximation is used to allow the fit for a continuous variation.
Additional fit details are discussed in the captions of Fig.~\ref{fig:g2_postfit} and Tab.~\ref{tab:fit_results}.

\begin{table}[h!]
\centering
\begin{tabular}{c|c|c}
Parameter & Best fit & Pull ($\sigma$)\\
\hline
\hline
$\eta_i$ & 0.319 $\pm$ 0.026 & -0.1 (3.9) \\
{$\eta_{s_1}$} & 0.370 $\pm$ 0.024 & 0.3 (2.5) \\{
$\eta_{s_2}$} & 0.436 $\pm$ 0.017 & 0.2 (3.3) \\
$\sigma_\text{CWDM}$ (nm) & 11.97 $\pm$ 0.95 & -1.0 (0.9) \\
$\sigma_p$ (GHz)& 87.7 $\pm$ 14.0 & 2.8 (1.4) \\
$k$ & $3.45 ^{+0.71}_{-0.50}$ & - \\
\hline
\end{tabular}
\caption{Maximum-likelihood best-fit results for key experimental parameters. 
Uncertainties are computed by inverting the Hessian, except for $k$, where a likelihood scan has been performed.
There is no value of pull for $k$ as it is extracted from a fit without a constraint.
}
\label{tab:fit_results}
\end{table}

\subsection{Single photon discrimination efficiency}
With $k=3.45^{+0.71}_{-0.50}$ extracted from the fit, the single-photon discrimination efficiency of our PNR detector is comparable to that of a pseudo-PNR detector comprised of approximately $ 11^{+7}_{-3}$ threshold detectors, each with efficiency $\eta_d=0.71$. 
Therefore, following the model developed in Sec.~\ref{sec:POVM_stats}, the experimental POVM is 
\begin{align}
    {\hat{\Pi}^{exp}_1} &\approx 0.710\ket{1}\bra{1}+0.458\ket{2}\bra{2}
    +0.222\ket{3}\bra{3}\\&+0.096\ket{4}\bra{4}+0.039\ket{5}\bra{5}+0.015\ket{6}\bra{6}\nonumber,
\end{align}
corresponding, according to the arguments in Sec.~\ref{sec:pnr_metric}, to a single photon discrimination efficiency of $\eta_{PNR}^1\approx0.46$, limited mainly by $\eta_d$.

\subsection{Improvement with a PNR SNSPD}
The reduction of $g^2(0)$ shown in Fig.~\ref{fig:g2_postfit} demonstrates a suppression of multi-photon events.
A maximum reduction of $\text{ }0.118\pm0.012$ at $\mu =0.327\pm0.007$ is observed; it is more clearly indicated in the inset of Fig.~\ref{fig:g2_postfit}.

\begin{figure}[h!]
    \centering
    \includegraphics[width=\columnwidth]{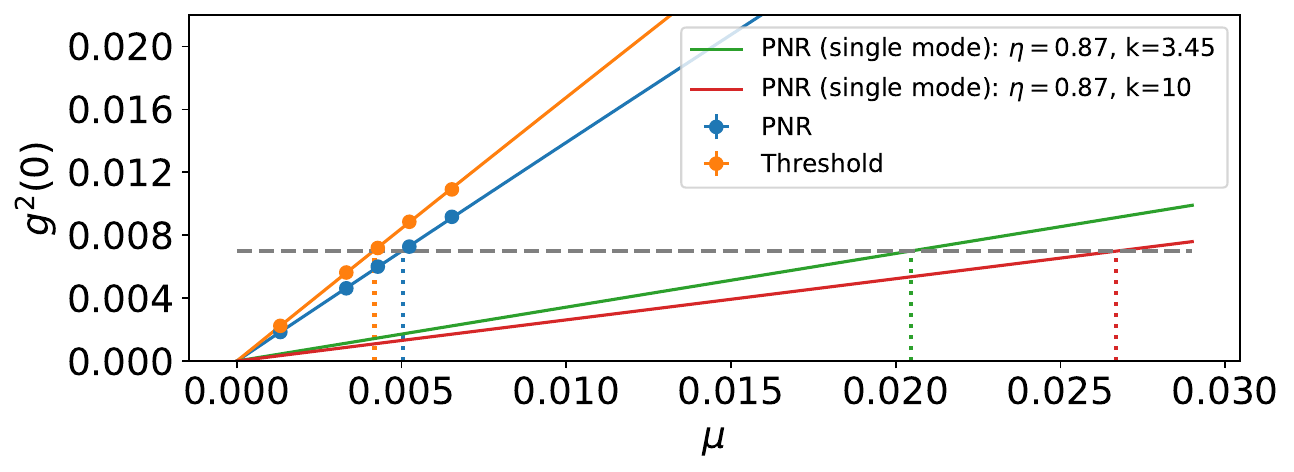}
    \caption{
    Correlation function $g^2(0)$ as a function of mean photon number $\mu<<1$ for our experiment and improved heralded single-photon sources. 
    The experimental data (large dots) are represented with their uncertainties.
    The models for the threshold configuration (orange curve) and PNR configuration (blue curve) of our detector are compared with the green and red curves, which correspond to model predictions using improved sources, as discussed in the main text, with key parameters (tree depth $k$, path efficiencies $\eta\equiv\eta_{s_1}=\eta_{s_2}=\eta_{i}$) shown in the inset. 
    Grey dashed line corresponding to a $g^2(0)$ measured in Ref. \cite{kaneda2019high}.
    }
    \label{fig:g2_postfit_zommed} 
\end{figure}

The data and fit for $\mu<<1$ is presented in Fig.~\ref{fig:g2_postfit_zommed}.
Configurations (ii) and (i) are denoted by orange and blue colors, respectively, with the data indicated by large dots and the fit by solid curves.
To give context, orange and blue dotted lines indicate the $\mu$ corresponding to a $g^2(0)$ of $7\times 10^{-3}$ (gray dashed line) measured in Ref. \cite{kaneda2019high}, Specifically, we observe a $25\%$ improvement in $\mu$, from $4\times 10^{-3}$ (orange dotted line) with configuration (ii), to $5\times 10^{-3}$ (blue dotted line) with configuration (i). 

To estimate the performance of our experiment with future improvements, we calculate $g^2(0)$ using the properties of our PNR detector ($k=3.45$, green curve) and those of a PNR detector with a higher tree depth ($k=10$, red curve).
We also assume higher path efficiencies of $\eta_{s_1}=\eta_{s_2}=\eta_{i}=0.87$, which are the product of the coupling (0.91) and detector (0.96) efficiencies from Refs.~\cite{kaneda2016heralded} and \cite{akhlaghi2015waveguide}, respectively, and are among the best-achieved to date. 
With these upgrades, for a $g^2(0)$ of $7\times 10^{-3}$ (gray dashed line), we predict an improved $\mu=20.5\times 10^{-3}$ (green curve) and $\mu = 26.7\times 10^{-3}$ (red curve) using our PNR SNSPD and a nearly ideal PNR detector, respectively.

\section{Discussion}
\label{sec:discussion}
By measuring the idler mode of a spontaneous parametric down-conversion source using a photon-number-resolving nanowire detector, we reduce the $g^2(0)$ of the signal mode or, on the other hand, increase the probability to generate a photon.
The results and key performance metrics of our experiment are supported by a detailed analytical model which captures multi-photon effects, imperfections, and multiple spectral modes.  Using a setup consisting of fiber-coupled and off-the-shelf devices, we generate photons that can be used in quantum information applications, in particular quantum communications \cite{ch1984quantum,krovi2016practical}.

To realize an ideal single photon source \cite{eisaman2011invited}, a number of improvements to our experiment must be implemented \cite{christ2012limits}.
First, the Schmidt number of our SPDC source must be decreased from its current value of $K\approx772$ to $K=1$. 
This can be accomplished by either narrower spectral filtering of the pairs or increasing the pump pulse bandwidth \cite{rarity1995interference}, the use of cavity-enhanced SPDC \cite{herzog1994frustrated}, or by engineering the phase matching function of the nonlinear crystal \cite{mosley2008heralded, xin2022}.
A near-unity Schmidt number renders the photons suitable for interference with other independently generated photons in a quantum circuit or network.

Next, the system efficiency should be increased to near unity.
Coupling between fibers and devices can be improved with enhanced modal engineering \cite{kawasaki1981biconical} or using anti-reflection-coated free-space components~\cite{shalm2015strong}. 
Alternatively, components could be integrated onto the same chip, for instance using Si- or SiN-on-insulator with SFWM sources~\cite{silverstone2014chip,wang2020integrated}, or using thin-film lithium niobate~\cite{zhu2021integrated}.
Furthermore, multiplexing strategies must be employed to increase the probability of generating a single pair beyond the theoretical maximum of 25\% per mode.
Such multiplexing, using, for instance, spatial~\cite{collins2013integrated,mendoza2016active}, temporal~\cite{xiong2016active,kaneda2016heralded}, or frequency modes~\cite{joshi2018frequency,puigibert2017heralded}, could also be employed to circumvent loss in the signal mode \cite{sinclair2014spectral}.
This requires on-demand feed-forward mode mapping using switches~\cite{xu2019low}, quantum memories~\cite{lvovsky2009optical}, or frequency shifters~\cite{hu2021chip}, respectively.
Feed-forward requires the real-time readout that our PNR SNSPD allows.
Note that feed-forward also allows for temporal filtering of the signal mode, a method that yields a significant reduction in $g^2(0)$ \cite{brida2012, krapick2013, massaro2019}.
We also point out that our improvement in $g^2(0)$ significantly reduces the number of spatially multiplexed sources ($\sim 1/\mu$ for $\mu<<1$) that are required to render our heralded photon source to be quasi-deterministic.
For instance, for $g^2(0)= 7\times10^{-3}$, in which we observe a 25\% improvement in $\mu$ from $4\times10^{-3}$ to $5\times10^{-3}$, see Fig. \ref{fig:g2_postfit_zommed}, corresponds to a reduction of the number of multiplexed spatial modes from 250 to 200.
Further, with an improved detector efficiency of 0.87 \cite{kaneda2016heralded,akhlaghi2015waveguide}, only $\sim 49$ multiplexed modes will be required to quasi-deterministically generate a heralded single photon.

Multiplexing with feed-forward also allows a multi-mode source to be rendered as single mode, i.e. it effectively decreases its Schmidt number to unity~\cite{puigibert2017heralded}.
Our broadband SPDC source is naturally suited for frequency multiplexing, as indicated by the strong frequency correlations in our JSI \cite{hiemstra2020pure}.
This suggests our measured $\mu=5\times10^{-3}$ for $g^2(0)= 7\times10^{-3}$ exceeds state-of-the-art SPDC sources using threshold detection, as well as quantum dots \cite{kaneda2019high}, accounting for such frequency multiplexing. 

Additional gains can be offered by improvements to the PNR SNSPD. 
A higher detector efficiency, i.e. ideally increasing $\eta_d$ to one, would increase the single-photon discrimination efficiency and improve the fidelity of the heralded single photon. 
This may be achieved through improvements to the optical stack around the nanowire by replacing the gold mirror with a distributed Bragg reflector mirror~\cite{reddy2020superconducting}.  Also, the detector reset time of nearly 100~ns restricts the maximum repetition rate of the source to be $\sim10$~MHz. An SNSPD with a reduced reset time based on a lower kinetic inductance nanowire material, or integrated with an active quenching circuit \cite{ravindran2020active}, would allow for high single-photon generation rates.  
A multiplexing method based on multiple PNR SNSPDs would also support a high repetition rate in addition to a substantial increase in detection efficiency \cite{bodog2020}.

Beyond single photon sources, extensions of our setup  allow efficient generation of qubits or qudits, as well as entanglement swapping using PNR SNSPDs~\cite{krovi2016practical}. Further uses encompass preparation of heralded photon-number states~\cite{cooper2013experimental} and non-Gaussian continuous-variable states~\cite{su2019conversion}, vital resources to realize fault-tolerant photonic quantum computers~\cite{bourassa2021blueprint}. Lastly, and of note, by using PNRs to improve teleportation rates \cite{valivarthi2020teleportation}, novel applications can benefit including microwave to optical transduction \cite{PhysRevLett.124.010511}. 
During the preparation of our manuscript we became aware of relevant results achieved independently of this work \cite{sempere2021reducing}.

\section{Acknowledgments} 
We acknowledge partial funding from the Department of Energy BES HEADS-QON Grant No. DE-SC0020376 (on applications related to transduction), QuantiSED SC0019219, and the AQT Intelligent Quantum Networks and Technologies (INQNET) research program.  
Partial support for this work was provided by the DARPA DSO DETECT, NASA SCaN, and Caltech/JPL PDRDF programs. 
S.I.D. and A.M. acknowledge partial support from the Brinson Foundation. 
Part of this research was performed at the Jet Propulsion Laboratory, California Institute of Technology, under contract with NASA.
We acknowledge productive discussions with Kayden Taylor, Sergio Escobar, Daniel Oblak, and Cristian Pe\~{n}a.  
We are grateful to Jason Trevor for technical assistance. 

\bibliography{main}

\end{document}